\documentclass[apj,onecolumn]{emulateapj}
\usepackage{amsmath}

\shortauthors{L. BOCO ET AL.}
\shorttitle{MERGING RATES OF COMPACT BINARIES IN GALAXIES}
\slugcomment{ACCEPTED BY APJ}

\begin{document}

\title{Merging Rates of Compact Binaries in Galaxies: \\ Perspectives for Gravitational Wave Detections}
\author{L. Boco\altaffilmark{$\star$,$\diamond$}, A. Lapi\altaffilmark{$\star$,$\diamond$,$\triangle$,$\Box$},\\ S. Goswami\altaffilmark{$\star$}, F. Perrotta\altaffilmark{$\star$}, C. Baccigalupi\altaffilmark{$\star$,$\diamond$}, L. Danese\altaffilmark{$\star$}}
\altaffiltext{$\star$}{SISSA, Via Bonomea 265, 34136 Trieste, Italy}\altaffiltext{$\diamond$}{IFPU, Via Beirut 2, 34014 Trieste, Italy}\altaffiltext{$\triangle$}{INFN-TS, via Valerio 2, 34127 Trieste,  Italy}\altaffiltext{$\Box$}{INAF-OATS, via Tiepolo 11, 34131 Trieste, Italy}

\begin{abstract}
We investigate the merging rates of compact binaries in galaxies, and the related detection rate of gravitational wave (GW) events with AdvLIGO/Virgo and with the Einstein Telescope. To this purpose, we rely on three basic ingredients: (i) the redshift-dependent galaxy statistics provided by the latest determination of the star formation rate functions from UV+far-IR/(sub)millimeter/radio data; (ii) star formation and chemical enrichment histories for individual galaxies, modeled on the basis of observations; (iii) compact remnant mass distribution and prescriptions for merging of compact binaries from stellar evolution simulations. We present results for the intrinsic birthrate of compact remnants, the merging rates of compact binaries, GW detection rates and GW counts, attempting to differentiate the outcomes among BH-BH, NS-NS, and BH-NS mergers, and to estimate their occurrence in disk and spheroidal host galaxies. We compare our approach with the one based on cosmic SFR density and cosmic metallicity, exploited by many literature studies; the merging rates from the two approaches are in agreement within the overall astrophysical uncertainties. We also investigate the effects of galaxy-scale strong gravitational lensing of GW in enhancing the rate of detectable events toward high-redshift. Finally, we discuss the contribution of undetected GW emission from compact binary mergers to the stochastic background.
\end{abstract}


\section{Introduction}\label{sec|introduction}

\setcounter{footnote}{0}

The recent detections of several gravitational wave (GW) events by the LIGO/Virgo collaborations (Abbott et al. 2016a,b,c; 2017a,b,c,d,e; 2019; also \texttt{https://www.ligo.org/}), and the many more expected with the advent of the upcoming advanced configurations and detectors like the Einstein Telescope (ET; see Sathyaprakash et al. 2012; Regimbau et al. 2012; also \texttt{http://www.et-gw.eu/}), are to provide tremendous breakthroughs in astrophysics, cosmology and fundamental physics (e.g., Taylor et al. 2012; Barack et al. 2018).

The GW events in the LIGO/Virgo operating frequency band are consistently interpreted as mergers of binary compact star remnants, e.g., neutron stars (NS) and/or black holes (BHs). On the one hand, the analysis of the individual GW signal waveforms can provide useful information about the properties and evolution of the progenitor binary systems (remnant masses, spins, orbital parameters; e.g., Weinstein 2012; Abbott et al. 2016a,b,c). On the other hand, the statistics of GW events can yield astrophysical constraints on stellar binary evolution (SN kicks, common envelope effects, mass transfers; e.g., Belczynski et al. 2016; Dvorkin et al. 2018; Mapelli \& Giocobbo 2018), on the average properties of the host galaxies (chemical evolution, star formation histories, initial mass function; e.g., O'Shaughnessy et al. 2010; de Mink \& Belczynski 2015; Vitale \& Farr 2018), and even on cosmology at large (e.g., Taylor \& Gair 2012; Nissanke et al. 2013; Liao et al. 2017; Fishbach et al. 2019).

In the present paper we will focus on forecasting the GW detection rate from merging compact binaries as a function of redshift, in the perspective of the next AdvLIGO/Virgo observing runs and of the future ET\footnote{Throughout the paper we refer to AdvLIGO/Virgo in the design configuration and to the ET in the ET-D xylophone configuration.}. The issue is complex because it involves numerous astrophysical processes occurring on vastly different time and spatial scales: from stellar astrophysics, to galaxy formation, to GW physics. A number of previous studies have approached the issue basing on population-synthesis simulations, that follow stellar and binary evolution so as to provide estimates of the remnant masses and merging timescales (e.g., Dominik et al. 2013, 2015; de Mink \& Belczynski 2015; Spera et al. 2015, 2019; Spera \& Mapelli 2017; Giacobbo \& Mapelli 2018). The compact binary merging rate has generally been derived by combining the above results with the cosmic star formation rate density and with a distribution of metallicity around the mean cosmic value, either inferred from observations (e.g., Belczynski et al. 2016; Lamberts et al. 2016; Cao et al. 2018; Elbert et al. 2018; Li et al. 2018) or derived from cosmological simulations (e.g., O'Shaughnessy et al. 2017; Mapelli et al. 2017; Lamberts et al. 2018; Mapelli \& Giacobbo 2018).

On the other hand, in the last decade a wealth of observations (e.g., UV+far-IR/sub-mm/radio luminosity functions and stellar/gas/dust mass functions, broadband spectral energy distribution, mass-metallicity relationships, size/kinematic evolution, etc.) have allowed to estimate the statistics of different galaxy populations as a function of their main physical properties across cosmic time; in addition, these observations have allowed to shed light on the age-dependent star formation and chemical enrichment histories of individual galaxies.

In this paper we will exploit these ingredients, in combination with the remnant mass distribution from single stellar evolution simulations (specifically, we rely on the \texttt{SEVN} code by Spera \& Mapelli 2017 based on the delayed SN engine and including pair-instability and pair-instability pulsation SNe, hereafter (P)PSNe) to compute GW detection rates in the perspective of the next AdvLIGO/Virgo observing runs and of the future ET detector. We also provide a tentative separation among the signals expected from BH-BH, NS-NS, and BH-NS merger events in disk and spheroidal galaxies.

The approach based on galaxy SFR and metallicity histories pursued here
can provide the joint probability distribution of chirp masses and host galaxy properties (SFR, stellar mass, metallicity, etc.) as a function of redshift (see also Belczynski et al. 2010a,b; O'Shaughnessy et al. 2010, 2017). We then predict how the detected GW event rates from high-redshift galaxies can be enhanced by strong galaxy-scale gravitational lensing. Finally, we investigate the contribution to the GW background expected from the incoherent superposition of undetected signals from compact binary mergers in galaxies.

The paper is organized as follows: in Sect.~\ref{sec|basics} we introduce the basic ingredients of our computation, including redshift-dependent galaxy statistics (see Sect.~\ref{sec|SFR_func}), star formation and chemical enrichment histories for individual galaxies (see Sect.~\ref{sec|SFR_hist}), and compact remnant mass distribution from stellar evolution simulations (see Sect.~\ref{sec|stellarevo}); then we compute compact remnant birthrates in Sect.~\ref{sec|birthrates} and intrinsic merging rates in Sect.~\ref{sec|mergerrates}; in Sect.~\ref{sec|GWdetection} we calculate the GW event detection rates expected in the next AdvLIGO/Virgo observing runs and the future ET detector, and in Sect.~\ref{sec|lensing} we discuss how these rates are affected by galaxy-scale gravitational lensing of GW; in Sect.~\ref{sec|GWback} we investigate the GW background from undetected events; finally, in Sect.~\ref{sec|summary} we summarize our findings.

Throughout this work, we adopt the standard flat $\Lambda$CDM cosmology
(Planck Collaboration 2019) with rounded parameter values: matter density $\Omega_M = 0.32$, baryon density $\Omega_b = 0.05$, Hubble constant $H_0 = 100\,h$ km s$^{-1}$ Mpc$^{-1}$  with $h = 0.67$, and mass variance $\sigma_8 = 0.81$ on a scale of $8\, h^{-1}$ Mpc. In addition, we use the widely adopted Chabrier (2003, 2005; see also Mo et al. 2010) initial mass function (IMF) with shape $\phi(\log m_\star)\propto \exp[-(\log m_\star-\log 0.2)^2/2\times 0.55^2]$ for $m_\star\la 1\, M_\odot$ and $\phi(\log m_\star)\propto m_\star^{-1.35}$ for $m_\star\ga 1\, M_\odot$, normalized as $\int_{0.08\, M_\odot}^{350\, M_\odot}{\rm d}m_\star\, m_\star\, \phi(m_\star)=1\, M_\odot$; the impact of the IMF choice on our results will be discussed in Sect. \ref{sec|GWdetection}. Finally, a value $Z_\odot\approx 0.015$ for the solar metallicity (Caffau et al. 2011) is adopted.

\section{Basic ingredients}\label{sec|basics}

Our analysis is based on three main ingredients: (i) an observational determination of the SFR function at different redshifts; (ii) average star formation and chemical enrichment histories of individual galaxies; (iii) outcomes from single stellar evolution simulations specifying the remnant masses for a given zero-age main sequence star. We now briefly present and discuss these in turn.

\subsection{SFR functions and cosmic SFR density}\label{sec|SFR_func}

The first ingredient is constituted by the SFR function ${\rm d}N/{\rm d}\log \psi\,{\rm d}V$, namely the number density of galaxies per comoving volume and per logarithmic bin of SFR $\psi$ at given cosmic time $t$ (corresponding to redshift $z$).

The SFR of a galaxy is directly proportional to the
intrinsic UV luminosity; however, the latter can be significantly absorbed by even a modest amount of dust and re-radiated mostly at far-IR/(sub)mm wavelengths (e.g., Kennicutt \& Evans 2012). For galaxies with relatively low SFRs $\psi\la 30-50\, M_{\odot}$ yr$^{-1}$ dust attenuation is mild and the intrinsic SFR can be soundly estimated from UV data alone via standard corrections based on the UV slope (see Meurer 1999; Calzetti et al. 2000; Bouwens et al. 2015). As a consequence, the SFR functions for SFRs $\psi\la 30-50\, M_{\odot}$ yr$^{-1}$ are rather well established by deep surveys in the rest-frame UV band, in some instances eased by gravitational lensing from foreground galaxy clusters, up to very high redshift $z\la 7-10$ (see Wyder et al. 2005; Oesch et al. 2010; van der Burg et al. 2010; Cucciati et al. 2012; Finkelstein et al. 2015; Alavi et al. 2016; Bouwens et al. 2016, 2017; Bhatawdekar et al. 2019; cf. open symbols in Fig.~\ref{fig|SFRfunc}).

On the other hand, in galaxies with high SFRs $\ga 30-50\, M_{\odot}$ yr$^{-1}$ dust absorption is heavier, and UV slope-based corrections are wildly dispersed and statistically fail (see Silva et al. 1998; Efstathiou et al. 2000; Coppin et al. 2015; Reddy et al. 2015; Fudamoto et al. 2017).
In this regime far-IR/(sub)mm observations becomes crucial to obtain sound estimates of the SFR; radio data can also be helpful, by eliciting the free-free emission associated with the ongoing SFR (e.g., Murphy et al 2012).
In fact, over recent years far-IR/(sub)mm wide-area surveys (see Lapi et al. 2011; Gruppioni et al. 2013, 2015, 2019; Magnelli et al. 2013; cf. filled symbols in Fig.~\ref{fig|SFRfunc}) have been exploited to reconstruct, in combination with the deep UV data mentioned above, the SFR functions of galaxies for redshifts $z\la 3$ over the whole range of relevant SFR $\psi\sim 10^{-2}$ to a few $10^3\, M_\odot$ yr$^{-1}$.

For redshifts $z\ga 3$ and large SFRs $\psi \ga 30-50\, M_{\odot}$ yr$^{-1}$ the shape of the SFR function is more uncertain, given the sensitivity limits of current wide-area far-IR surveys. However, relevant constraints have been obtained recently from deep radio surveys (Novak et al. 2017), from far-IR/(sub)mm stacking (see Rowan-Robinson et al. 2016; Dunlop et al. 2017) and super-deblending (see Liu et al. 2018) techniques, and from targeted far-IR/(sub)mm observations of significant yet not complete samples of starforming galaxies (e.g., Riechers et al. 2017; Marrone et al. 2018; Zavala et al. 2018) and quasar hosts (e.g., Venemans et al. 2017; Stacey et al. 2018).

The resulting SFR functions at representative redshifts are illustrated in Fig.~\ref{fig|SFRfunc}. These can be smoothly rendered, over the redshift range $z\sim 0-8$ for SFR $\psi\sim 10^{-2}$ to a few $10^3\, M_\odot$ yr$^{-1}$, with a simple Schechter shape
\begin{equation}
{{\rm d}N\over {\rm d}\log\psi\,{\rm d}V}(\psi,t) = \mathcal{N}(z)\, \left[{\psi\over \psi_c(z)}\right]^{1-\alpha(z)}\,e^{-\psi/\psi_c(z)}~,
\label{eq|SFRfunc}
\end{equation}
in terms of three redshift-dependent parameters $\mathcal{N}(z)$, $\alpha(z)$, $\psi_c(z)$, as specified in Mancuso et al. (2016b; see their Table 1). In Mancuso et al. (2016a,b; 2017) and Lapi et al. (2017a,b) the SFR functions have been validated against independent datasets, including integrated galaxy number counts at relevant far-IR/(sub)mm/radio wavelengths, counts/redshift distributions of strongly gravitationally-lensed galaxies,  main sequence of star-forming galaxies. An additional, straight test for the SFR functions, performed by Lapi et al. (2017b; see their Fig.~4), is the computation of the stellar mass function via the continuity equation, directly connecting the star formation to the building up of the stellar mass in galaxies, and the comparison with statistical observations at different redshifts for both quiescent and starforming objects (e.g., Davidzon et al. 2017). At $z\ga 1$ the bright end of the SFR function turns out to be populated by heavily dust obscured, strongly starforming galaxies, which  constitute the progenitors of local massive spheroids with masses $M_\star\ga$ a few $10^{10}\, M_\odot$; the faint end is instead mainly populated by mildly obscured starforming galaxies, that will end up in spheroid-like objects with stellar masses $M_\star\la 10^{10}\, M_\odot$. On the other hand, disk-dominated galaxies with stellar masses $M_\star\la$ several $10^{10}\, M_\odot$ are found to be well traced by the UV-inferred SFR function at $z\la 2$.

From the SFR functions, the cosmic SFR density (per unit comoving volume) is straightforwardly computed as
\begin{equation}
\rho_{\psi}(z) = \int{\rm d}\log \psi\, {{\rm d}N\over {\rm d}\log \psi\,{\rm d}V}\, \psi~.
\label{eq|cosmicSFR}
\end{equation}
The outcome is illustrated in Fig.~\ref{fig|SFRcosm} (black solid line) and compared with available multi-wavelength datasets; the literature estimates from dust-corrected UV observations by Madau \& Dickinson (2014), from SNI$a$ searches at high redshift by Strolger et al. (2004), and from high-redshift long GRBs by Kistler et al. (2013) are also reported for reference. Notice that the cosmic SFR density constructed from the latest determination of the SFR functions (see Fig.~\ref{fig|SFRfunc}) is appreciably higher than previous estimates, and peaks toward slightly higher redshift; this can be traced back to a more complete sampling of the dusty starforming galaxy population for $z\ga 2$ thanks to the most recent wide-area far-IR/(sub)mm/radio surveys (see Gruppioni et al. 2013, 2019; Rowan-Robinson et al. 2016; Novak et al. 2017; Liu et al. 2018).

\subsection{Star-formation and metal-enrichment history of individual galaxies}\label{sec|SFR_hist}

The second ingredient of our analysis is constituted by the history of star formation and chemical enrichment in individual galaxies. The quantities relevant for the present paper are the average behaviors of the SFR $\psi(\tau)$ and of the global metallicity $Z(\tau)$ as a function of the internal galactic age $\tau$ (i.e., the time since the beginning of significant star formation activity) for a galaxy observed at cosmological time $t$.

As to the star-formation history, for high $z\ga 2$ starforming galaxies many SED-modeling studies (e.g., Papovich et al. 2011; Smit et al. 2012; Moustakas et al. 2013; Steinhardt et al. 2014; Citro et al. 2016; Cassar\'a et al. 2016) suggest to describe the star formation history as a truncated power-law shape
\begin{equation}
\psi(\tau) \propto \tau^\kappa\, \Theta_{\rm H}(\tau-\tau_\psi)~,
\label{eq|SFRhist}
\end{equation}
where $\kappa\la 0.5$ controls the slow power-law rise, and $\Theta_{\rm H}(\cdot)$ is the Heaviside function specifying the star formation duration $\tau_{\psi}$. The latter can be inferred from the galaxy main sequence (see Daddi et al. 2007; Rodighiero et al. 2011, 2015; Speagle et al. 2014, Dunlop et al. 2017), a well-known relationship linking the peak value of the SFR $\psi(\tau_\psi)$ to the relic stellar mass $M_\star(\tau_\psi)= \int_0^{\tau_\psi}{\rm d}\tau\, \psi(\tau)$; specifically, the redshift-dependent main sequence relation measured via multi-wavelength data by Speagle et al. (2014) is used to compute $\tau_\psi$. This yields a star formation duration of $\la$ Gyr for strongly starforming objects with $\psi\ga 10^2\, M_\odot$ yr$^{-1}$, which are the progenitors of massive spheroids with $M_\star\ga$ a few $10^{10}\, M_\odot$. Such a short timescale is also in line with local observations of the $\alpha-$enhancement in massive early-type galaxies; this represents an iron underabundance compared to $\alpha$ elements, that occurs because star formation is stopped, presumably by some form of energetic feedback (e.g., due to the central supermassive black hole), before type I$a$ SN explosions can pollute the interstellar medium with substantial iron amounts (e.g., Romano et al. 2002; Gallazzi et al. 2006; Thomas et al. 2005, 2010; Johansson et al. 2012). Conversely, in low-mass spheroidal galaxies with $M_\star\la 10^{10}\, M_\odot$ the star formation durations $\tau_\psi$ inferred from the main sequence are much longer, amounting to a few Gyr as also indicated by data on the age of stellar population and on chemical abundances (see review by Conroy 2013). Finally, in low redshift $z\la 2$ disk-dominated galaxies classic evidence indicates that on average the star formation rate declines exponentially $\psi(\tau)\propto e^{-\tau/\tau_\psi}$ over a long characteristic timescale $\tau_\psi\approx$ several Gyrs (see Chiappini et al. 1997; Courteau et al. 2014; Pezzulli \& Fraternali 2016; Grisoni et al. 2017). Lapi et al. (2017a, 2018) have shown that the above star formation histories can be exploited to connect, via the continuity equation, the SFR functions to the observed stellar mass functions at different redshifts (e.g., Davidzon et al. 2017), for both starforming and quiescent galaxies.

We caveat that the aforementioned star formation histories for spheroids and disks are meant to represent the average statistical behavior of the respective galaxy population, and to render for each galaxy the spatial and time integrated behaviors. This is clearly an approximation that may be not realistic in specific objects and/or around particular spatial locations. As an example, galaxies featuring multiple recurrent bursts of star formation may be preferential hosts of double compact objects mergers, especially if short time delays between the birth and the coalescence of the compact binaries are favored (see Sect.~\ref{sec|mergerrates}). As another example, in the Milky Way local constraints from observations of the solar neighborhood (e.g., Cignoni et al. 2006) seems not to favor an exponentially declining SFR but rather to suggest a low-level constant value with an enhancement around $3$ Gyr ago, although these findings are still somewhat debated (e.g., Bovy et al. 2017).

As to the chemical enrichment history of individual galaxies, we have exploited the standard code \texttt{che-evo} incorporated into \texttt{GRASIL} (Silva et al. 1998, 2011; Bressan et al. 2002; Panuzzo et al. 2003; Vega et al. 2005).
For spheroidal galaxies, it reproduces the observed local relationship between stellar metallicity and stellar mass, and its weak evolution with redshift (see Arrigoni et al. 2010; Spolaor et al. 2010; Gallazzi et al. 2014). For disk galaxies at $z\la 2$, it reproduces the observed relationship between gas metallicity and stellar mass, including its appreciable redshift-dependence (see Andrews \& Martini 2013; Zahid et al. 2014; de la Rosa et al. 2015; Onodera et al. 2016). In both cases, within an individual galaxy the metallicity behavior is closely approximated by
\begin{equation}
Z(\tau)\simeq
\left\{
\begin{aligned}
&Z_{\rm sat}\,{\tau\over \Delta\tau_\psi}~~~~~~~~  &{\tau\over \tau_\psi}\leq \Delta\\
\\
&Z_{\rm sat} ~~~~~~~~ &{\tau\over \tau_\psi}\geq \Delta
\end{aligned}
\right.
\label{eq|metalevolution}
\end{equation}
i.e., it increases from $Z=0$ almost linearly with the galactic age, and then after a time $\tau=\Delta\, \tau_\psi$, saturates to the value $Z_{\rm sat}$. The dependence of $Z_{\rm sat}$ and $\Delta$ on SFR/stellar mass can be described with an expression inspired by analytic chemical evolution models (see Cai et al. 2013; Feldmann 2015); this yields
\begin{equation}
\left\{
\begin{aligned}
&Z_{\rm sat}\propto {s\, y_Z\, (1-\mathcal{R})\over s\,(1-\mathcal{R}+\epsilon_{\rm out})-1}\\
&\\
&\Delta\simeq {1\over 3\,(1-\mathcal{R}+\epsilon_{\rm out})}\\
\end{aligned}
\right.~~~~~~~\epsilon_{\rm out}\approx 2\, \left(M_\star\over 10^{10}\, M_\odot\right)^{-0.25}~;
\label{eq|chemidetail}
\end{equation}
here $s\approx 3$ is the ratio between the dynamical timescale of the infalling gas and the star formation timescale, $\mathcal{R}\approx 0.44$ is the recycling stellar mass fraction, $y_Z\,(1-\mathcal{R})\approx 0.034$ is the metal yield (assuming the Romano et al. 2010 stellar yields), and $\epsilon_{\rm out}$ is the mass loading factor of galactic outflows from stellar winds and supernova explosions. In the above equation we have provided a fit for $\epsilon_{\rm out}$ as a function of the final stellar mass $M_\star$ from the results of \texttt{che-evo} code; a similar expression concurrently describes the time-averaged outcome from hydrodynamical simulations of stellar feedback (e.g., Hopkins et al. 2012).
As a result, typical values $Z_{\rm sat}\sim 0.3-1.5\, Z_{\odot}$ are obtained for galaxies with final stellar masses in the range $M_{\star}\sim 10^{9-11}\, M_\odot$, respectively (see, e.g., Chruslinska et al. 2019); the related quantity $\Delta\sim 0.1-0.3$ specifies how quickly the metallicity saturates to such values as a consequence of the interplay between cooling, dilution, and feedback processes. Note that several chemical evolution codes present in the literature, reproducing comparably well observations on the chemical abundances in galaxies of different stellar masses, also share a similar age-dependent metallicity behavior.

In this paper we will exploit the above gas metallicity evolution of individual galaxies as a function of time and SFR/stellar mass to compute merging rates of compact remnants and related GW event detection rates. The metallicity evolution enters into play since the mass distribution of the compact remnants depends on the chemical composition of the star-forming gas (see Sect.~\ref{sec|stellarevo}). In previous works an alternative, simpler approach has often been adopted, that involves the use of the mean cosmic metallicity (cf. Madau \& Dickinson 2014)
\begin{equation}
\langle Z(z)\rangle \approx {y_Z\,(1-\mathcal{R})\over \rho_b}\int_z{\rm dz'}\,\rho_{\psi}(z')\, \left|{{\rm d}t\over {\rm d}z'}\right|~,
\label{eq|cosmicmetallicity}
\end{equation}
where $\rho_b\approx 2.8\times 10^{11}\, \Omega_b\, h^2\, M_\odot$ Mpc$^{-3}$ is the background baryon density, and $\rho_\psi(z)$ is the cosmic SFR density. We report as a thin line in the inset of Fig.~\ref{fig|SFRcosm} the result of this procedure (solid black line). The outcome turns out to be consistent with measurements of the IGM metallicity as inferred from Ly$\alpha$ forest absorption lines (e.g., Aguirre et al. 2008), while it falls short with respect to the metallicity of damped Ly$\alpha$ absorption systems (e.g., Rafelski et al. 2012) and to the metal abundances in the central regions of galaxy clusters (e.g., Balestra et al. 2007). This is why in previous works on merging rates (e.g., Belczynski et al. 2016; Cao et al. 2018), a floor value of $0.5$ dex in $\log\langle Z(z)\rangle$ has been added to better fit such observational data (thick line in the inset of Fig.~\ref{fig|SFRcosm}; see also Vangioni et al. 2015); moreover, a log-normal distribution of metallicity around this mean cosmic value with a $1\sigma$ dispersion $\sigma_{\log Z} = 0.5$ dex has been also usually adopted. Note that even after such renormalization and scatter, the  cosmic metallicity stays appreciably lower than the saturation value of the gas metallicity in individual star-forming galaxies (see also Chruslinska et al. 2019). In the sequel, we will present results from the cosmic metallicity approach for comparison with our findings based on the metallicity evolution in individual galaxies.

\subsection{Remnant mass distribution from stellar evolution simulations}\label{sec|stellarevo}

We adopt the metallicity-dependent relationships $m_\bullet(m_\star,Z)$ between compact remnant mass $m_\bullet$ and zero-age main sequence star mass $m_\star$ provided by Spera \& Mapelli (2017). These have been obtained via specific simulations of single stellar population synthesis with the code \texttt{SEVN}, which couples the \texttt{PARSEC} stellar evolution tracks up to very massive stars (see Bressan et al. 2012; Tang et al. 2014; Chen et al. 2015) with up-to-date recipes for SN explosions (see Fryer et al. 2012). In particular, as a default we adopt their model based on the delayed SN engine, and including (P)PSNe (see also Woosley 2017). The mass $m_\bullet(m_\star,Z)$ of compact remnants is illustrated in Fig.~\ref{fig|remnant} for different metallicities. This has been obtained by interpolating on fine grids in $m_\star$ and $Z$ the tabulated data provided by Spera \& Mapelli (2017). Fig.~\ref{fig|remnant} can also be helpful for the reader to recognize how our results presented in next Sections will depend on detailed features of the remnant mass distribution as a function of metallicity.

To take into account modeling uncertainties and physical spread related mainly to stellar evolution processes like mass loss, SN mechanism, rotation/mixing, pulsations, etc.), we describe the mass distribution ${\rm d}p/{\rm d}\log m_\bullet$ of compact remnants with a log-normal function centered on the Spera et al. (2017) relation $m_\bullet(m_\star,Z)$, adopting a $1\sigma$ variance of $\sigma_{\log m_\bullet}=0.1$ dex:
\begin{equation}
{{\rm d}p\over {\rm d}\log m_\bullet}(m_\bullet|m_\star,Z) = {1\over \sqrt{2\pi}\,\sigma_{\log m_\bullet}}\, e^{-[\log m_\bullet -\log m_\bullet(m_\star,Z)]^2/2\,\sigma_{\log m_\bullet}^2}~.
\label{eq|remnant}
\end{equation}
We caveat the reader that the average relation $m_\bullet(m_\star,Z)$
by Spera et al. (2017) does not include binary evolutionary effects (e.g., mass transfers, common envelope and stellar mergers, tidal evolution, etc.), although incorporating these in the \texttt{SEVN} code yields a
remnant mass distribution not significantly different from the one from single stellar evolution (see Spera et al. 2019).

\section{Birthrates of compact remnants in galaxies}\label{sec|birthrates}

We start by computing the birthrate for a remnant mass $m_\bullet$ at cosmic time $t$ per unit comoving volume. This can be written as
\begin{equation}
R_{\rm birth}(m_\bullet,t) \simeq \int{\rm d}\log\psi\,\psi\,{{\rm d}N\over {\rm d}\log\psi\, {\rm d}V}(\psi,t)\,\int{\rm d}\log Z\, {{\rm d}p\over {\rm d}\log Z}(Z|\psi,t)\, \int {\rm d} m_\star \phi(m_{\star})\,{{\rm d} p\over {\rm d} m_\bullet}(m_\bullet|m_\star, Z)~
\label{eq|easybirthrate}
\end{equation}
The rationale behind this expression is the following. In the inner integral the mass distribution of compact remnants ${\rm d}p/{\rm d}\log m_\bullet$ from Eq.~(\ref{eq|remnant}), dependent on star mass and metallicity, is weighted with the IMF\footnote{Note that in Eq.~(\ref{eq|easybirthrate}) and following ones, the inner integral over star masses should contain the quantity $\phi(m_\star)/\int{\rm d}m_\star\, \phi(m_\star)\, m_\star$; however, in the literature the denominator is usually left implicit because of the IMF normalization condition $\int{\rm d}m_\star\, \phi(m_\star)\, m_\star\equiv 1\, M_\odot$, though the reader should keep track of the measure units.} $\phi(m_\star)$; the minimum star mass originating a NS remnant is set to $7\, M_\odot$. Then the outcome is averaged over the metallicity distribution, dependent on SFR and cosmic time, and then weighted by the SFR of the galaxy and the related galaxy number densities (namely, the SFR functions).

The metallicity distribution is in turn derived from the metallicity evolution within individual galaxies as expressed by Eq.~(\ref{eq|metalevolution}), taking into account the fractional time spent by the galaxy in a given metallicity bin
\begin{equation}
{{\rm d} p\over {\rm d} \log{Z}}\approx \Delta\times {Z\over Z_{\rm sat}}\, \ln(10)\, \Theta_{\rm H}(Z-Z_{\rm sat})+(1-\Delta)\times \delta_D[\log Z-\log Z_{\rm sat}]~,
\label{eq|metaldist}
\end{equation}
where $\delta_D[\cdot]$ is the Dirac-delta function and $\Delta$ is provided by the chemical evolution code as a function of SFR and redshift.
In our approach galaxies with the same final stellar mass picked up at the same redshift and galactic age would feature the same gas metallicity; however, the above metallicity distribution originates since a galaxy of given final mass, observed at redshift $z$ can have different ages depending on its formation redshift. Actually we convolve the above distribution with a Gaussian kernel featuring a dispersion of $0.15$ dex, that corresponds to the scatter estimated for the ISM metallicity in galaxies at given SFR but varying stellar mass (e.g., Mannucci et al. 2010; Salim et al. 2015; Sanders et al. 2018). To sum up, in each galaxy the metallicity $Z$ increases linearly with galactic age (cf. Eq.~\ref{eq|metalevolution}), so that stars born at different times have different $Z$; this creates an ever-changing
distribution of metallicities in each individual galaxy, that depends
on galaxy birthtime and mass.

We note that very often in the literature the detailed star formation and chemical enrichment history of individual galaxies are neglected, and an approach based on the cosmic SFR density and cosmic metallicity from Eqs.~(\ref{eq|cosmicSFR}) and (\ref{eq|cosmicmetallicity}) is adopted instead; in these studies the metallicity distribution ${\rm d} p/{\rm d} \log{Z}$ is taken to be a broad log-normal function with a $1\sigma$ dispersion of $\sigma_{\log Z}\approx 0.5$ dex around the mean cosmic value $\langle Z(z)\rangle$ of Eq.~\ref{eq|cosmicmetallicity} (including the $0.5$ dex increase, see Sect.~\ref{sec|SFR_hist}). In such a case, in Eq.~(\ref{eq|easybirthrate}) the dependence on the galaxy SFR is eliminated, so that the outermost integral yields simply the cosmic SFR density and the birthrate is given by\footnote{Actually $\rho_{\psi}$ and $\langle Z\rangle$ should be computed at $t-\tau_{m_\star}$ where $\tau_{m_\star}$ is the star lifetime; however, since $\tau_{m_\star}\ll t$ this delay is safely neglected.}
\begin{eqnarray}
R_{\rm birth}(m_{\bullet}, t)\simeq \rho_{\psi}(t)\int {\rm d}\log Z\,{{\rm d} p\over {\rm d} \log Z}(Z|\langle Z\rangle[t])\int {\rm d} m_\star \phi(m_{\star})\,{{\rm d} p\over {\rm d} m_\bullet}(m_\bullet|m_\star, \langle Z\rangle[t]).
\label{eq|cosmicbirthrate}
\end{eqnarray}

In Fig.~\ref{fig|Rbirth} we illustrate the birthrate $R_{\rm birth}(m_\bullet,t)$ for different redshifts; in particular, black lines refer to our computation based on Eq.~(\ref{eq|easybirthrate}) taking into account the star formation and chemical enrichment histories of individual galaxies, blue lines highlight the contribution in low $z\la 2$ disks, while green lines show the result based on Eq.~(\ref{eq|cosmicbirthrate}) relying on the cosmic SFR density and cosmic metallicity. One can recognize three characteristic features in these curves. First, there is a prominent NS peak at around $m_\bullet\sim 1.4\, M_\odot$; this reflects the higher birthrate of NS with respect to BH by a factor of $2-3$, as expected given the bottom-heavy shape of the adopted Chabrier IMF. Second, there is a BH plateau for $m_\bullet\sim 2.5-25\, M_\odot$, produced by the steep increasing shape of the Spera et al. (2017) relation $m_\bullet(m_\star|Z)$ in that range, that offsets the IMF decline. Third, $R_{\rm birth}$ falls off for high remnant masses $m_\bullet\ga 30\, M_\odot$, due to the behavior of the Spera et al. (2017) relation (see Sect.~\ref{sec|SFR_hist}). As to the redshift dependence, it mainly reflects the behavior of the SFR function (or of the cosmic SFR density), which increases out to $z\approx 2.5$ (solid line in Fig.~\ref{fig|SFRcosm}), and then recedes for higher redshift.

The comparison between our computation taking into account the star formation and metal enrichment histories of individual galaxies, and the approach based on the cosmic SFR density and cosmic metallicity is easily understood. At low redshift, the birthrate shape between the two approaches is similar, since the cosmic metallicity values are close to those applying in galaxies. In moving toward higher redshifts the cosmic metallicity is on average lower than that in individual galaxies, causing an enhancement in the relative occurrence of more massive BHs. In the birthrate computed with the cosmic approach, this causes a lower BH plateau in the range $m_\bullet\approx 2.5-25\, M_\odot$, followed by a peak at around $m_\bullet\approx 25-50\, M_\odot$, and then a more extended tail toward higher masses.

\section{Merging rates of compact remnants in galaxies}\label{sec|mergerrates}

The merging rate per unit volume and mass of the primary (more massive) remnant $m_\bullet$ is given by
\begin{equation}
R_{\rm merge}(m_{\bullet}, t)=f_{\rm eff}\int_{t_{d,{\rm min}}}^t {\rm d} t_d\, {{\rm d} p\over {\rm d} t_d}(t_d)\, R_{\rm birth}(m_{\bullet}, t-t_d)~,
\label{eq|mergratetemp}
\end{equation}
where $t_d$ is the delay time between the formation of the compact binary system and the merging event; a number of independent studies based on observations (see review by Maoz et al. 2014) and simulations (e.g., Dominik et al. 2012; Giacobbo \& Mapelli 2018) suggest a delay time probability distribution with shape ${\rm d}p/{\rm d}t_d\propto t_{d}^{-1}$, normalized to unity between a minimum value $t_{d,{\rm min}}\approx 50$ Myr and the age of the Universe, independently of the compact binary type involved. The factor $f_{\rm eff}$ in Eq.~(\ref{eq|mergratetemp}) is defined as the fraction of primary compact remnants that are hosted in binary systems with characteristics apt to allow merging of the companions within a Hubble time; it will be discussed below.

We are now in position to compute the merging rate of compact binaries
as a function of redshift by integrating out $R_{\rm merge}(m_{\bullet}, t)$ with respect to the compact remnant mass. The chirp mass $\mathcal{M}_{\bullet\bullet}$ and the primary remnant mass $m_\bullet$ can be related by $\mathcal{M}_{\bullet\bullet}=m_{\bullet}\, q^{3/5}/(1+q)^{1/5}$ where $q$ is the mass ratio between the companion and the primary remnant. Introducing a mass ratio distribution ${\rm d}p/{\rm d}q$ and changing variable from $m_{\bullet}$ to $\mathcal{M}_{\bullet\bullet}$ via a simple jacobian, one obtains
\begin{equation}
R_{\rm merge}(t)=\int{\rm d}\mathcal{M}_{\bullet\bullet}\,R_{\rm merge}(\mathcal{M}_{\bullet\bullet},t)=\int{\rm d}\mathcal{M}_{\bullet\bullet}\, \int{\rm d}q\, {{\rm d} p\over {\rm d} q}(q)\,{(1+q)^{1/5}\over q^{3/5}}\,
R_{\rm merge}\left[\mathcal{M}_{\bullet\bullet}\,{(1+q)^{1/5}\over q^{3/5}}, t\right]~.
\label{eq|mergerate}
\end{equation}
This expression is general, and in the following we will exploit it to estimate the merging rates for BH-BH, NS-NS and BH-NS by inserting the appropriate $q-$distribution and integration limits.

Specifically, we take the mass ratio distribution for BH-BH mergers to be ${\rm d} p/{\rm d} q\propto q$ with a minimum value $q_{\rm min}=0.5$; this yields an average mass ratio $\langle q\rangle \approx 0.8$, as suggested by stellar evolution simulations (see de Mink et al. 2013; Belczynski et al. 2016). On the other hand, for systems like NS-NS or BH-NS the shape of the mass-ratio distribution can be different (see Dominik et al. 2012, 2015; de Mink \& Belczynski 2015; Chruslinska et al. 2018; Mapelli \& Giacobbo 2018); specifically, it is found that low values $q<0.5$ apply for most BH-NS events and values $q\la 1$ for most of the NS-NS events ($q\approx 0.7-1$ for the GW170817 event, see Abbott et al. 2017c). On this basis, we assume a flat distribution for BH-NS in the range $q\sim 0-0.5$ and for NS-NS mergers in the range $q\sim 0.4-1$; we have also checked that the overall results for the merging rate depend very little on the adopted mass-ratio distributions. Note that the integrand $R_{\rm merg}(\mathcal{M}_{\bullet\bullet},t)$ in Eq.~(\ref{eq|mergerate}) represents the merging rate per unit chirp mass and will be extensively used below when dealing with GW event detection.
We separate the merging rate for BH-BH, NS-NS and BH-NS events basing on the primary mass: if it is in the range $1-2.5\, M_\odot$ the event is considered a NS-NS merger, otherwise we consider the appropriate limits in the mass-ratio distribution to obtain BH-NS and BH-BH rates.

Coming back to the parameter $f_{\rm eff}$, we caveat that it is the results of many different and complex physical processes related to stellar and dynamical evolution (e.g., binary fraction, common envelope development/survival, SN kicks, mass transfers, etc.). In ab-initio stellar evolution simulations (possibly including binary effects; e.g., O' Shaughnessy et al. 2010; Dominik et al. 2015; Belczinski et al. 2016; Spera et al. 2017; Giacobbo \& Mapelli 2018) this quantity is naturally obtained, though the outcomes are somewhat dependent on model assumptions. Alternatively, it can be set empirically by normalizing the local BH-BH merging rates (e.g., Dvorkin et al. 2016; Cao et al. 2018; Li et al. 2018) to the measurements by LIGO/Virgo (see Abbott et al. 2016c, 2017e, 2019); specifically, here we normalize the local BH-BH rate to the average logarithmic value $30$ Gpc$^{-3}$ yr$^{-1}$ from the latest interval estimation $9.7-101$ Gpc$^{-3}$ yr$^{-1}$ by LIGO/Virgo (see Fig. 12 in Abbott et al. 2019; cf. cyan shaded area in Fig.~\ref{fig|Rmerg}), implying $f_{\rm eff}\approx 2\times 10^{-4}$. We caveat that such a normalization is meaningful as long as the
local BH-BH merger rate detected by LIGO can be traced back to the binary compact remnants considered in the present paper, and it is not substantially contributed by additional channels (e.g., primordial BHs, globular/open cluster BHs, pop-III star BHs, etc.).

Note, however, that in principle $f_{\rm eff}$ could depend on the remnant masses and/or binary type. In the latter case, one could in principle set $f_{\rm eff}$ for NS-NS and for BH-NS events still relying on estimates of the local merging rate from various observations; however, these are very uncertain and somewhat in tension with each other. For example, local NS-NS rates $R_{\rm merge,NS-NS}\approx 110-3840$ Gpc$^{-3}$ yr$^{-1}$ are estimated by LIGO from the unique event GW170817 detected in GW so far (see Abbott et al. 2017b,c, 2019; cf. orange shaded area in Fig.~\ref{fig|Rmerg}). Chruslinska et al. (2018), using as input the Galactic merging rate $\approx 21^{+28}_{-14}$ Myr$^{-1}$ from observations of double pulsars (see Kim et al. 2015), obtain a low local NS-NS merging rate $R_{\rm merge, NS-NS}\approx 50$ Gpc$^{-3}$ yr$^{-1}$; these authors point out that, with a specific parameter choice in their models, this rate can be enhanced up to $R_{\rm merge, NS-NS}\approx 600^{+600}_{-300}$ Gpc$^{-3}$ yr$^{-1}$ but at the cost of overestimating the local BH-BH rate. Della Valle et al. (2018; see also Jin et al. 2018 and Pol et al. 2019) estimate a local NS-NS rate $R_{\rm merge, NS-NS}\approx 352^{+810}_{-281}$ Gpc$^{-3}$ yr$^{-1}$ for short GRB/kilonova events similar to GRB170817A, that can be made more consistent with the LIGO result by assuming a rather large viewing angle distribution for the beamed emission. NS-NS merging rates can also be inferred from the galactic abundance of elements produced via rapid neutron capture processes (in particular, Europium), but precise estimates are hindered by large uncertainties in chemical yields (e.g., Cote et al. 2018). As for the rate of BH-NS mergers only an upper limit is available $R_{\rm merge,BH-NS}\la 610$ Gpc$^{-3}$ yr$^{-1}$ (see Abbott et al. 2016d, 2019).

Given the current large theoretical and observational uncertainties, in the following we assume the aforementioned value of $f_{\rm eff}$ based on the local BH-BH merging rate as a reference also for both NS-NS and BH-NS events. Correspondingly, this yields local rates $R_{\rm merge, NS-NS}\approx 70$ Gpc$^{-3}$ yr$^{-1}$ and $R_{\rm merge, BH-NS}\approx 20$ Gpc$^{-3}$ yr$^{-1}$, respectively; the NS-NS rate so obtained is smaller than the LIGO estimate, and more consistent with those from double pulsars in the Milky Way, while short GRBs/kilonova occurrence are in between. More statistics from GW detectors are needed before drawing definite conclusions about the difference between $f_{\rm eff}$ for BH-BH and NS-NS or BH-NS mergers; however, if the high LIGO/Virgo local NS-NS rate were confirmed, a consequence would be that binary effects may work diversely for binary NS with respect to binary BH progenitors (see Mapelli \& Giacobbo 2018), leading in turn to an appreciable difference in the respective factors $f_{\rm eff}$. In that case, our results for NS-NS and BH-NS rates as a function of redshift can be simply rescaled by an overall normalization.

Our results for the merging rates as a function of redshift are illustrated in Fig.~\ref{fig|Rmerg}. The cyan (orange) shaded area is the LIGO measurement of the BH-BH (NS-NS) merging rate at $z\approx 0$. Solid black lines illustrate the merging rates for BH-BH events, dashed for NS-NS and dotted for BH-NS. It can be seen that the merging rate for NS-NS events is appreciably higher than for BH-BH; this reflects the corresponding behavior of the compact remnant birthrate $R_{\rm birth}$, cf. Fig.~\ref{fig|Rbirth}. For the adopted Chabrier IMF, most of the stars evolving into a compact remnant become NSs rather than BHs; thus an intrinsically larger merging rate for NS-NS than for BH-BH is naturally originated. However, we anticipate that such a ratio between NS-NS and BH-BH mergers is no longer valid when considering GW detectable events, due to the dependence of the GW signal on the chirp/total mass of the binary (cf. Sect.~\ref{sec|GWdetection}). Note that the contribution from disk-dominated galaxies to the overall merging rate is subdominant. In particular, at $z\approx 0$ events hosted in disk galaxies contribute around $25\%$ to the overall rate.

In the same Fig.~\ref{fig|Rmerg} black lines refer to our approach taking into account star formation and chemical evolution histories of individual galaxies, while green lines refer to the computation based on the cosmic SFR density and cosmic metallicity. The overall intrinsic merging rates $R_{\rm merge}(t)$ in the two approaches (cf. top panel in Fig.~\ref{fig|Rmerg}) are in agreement. 
In fact, when integrating over the chirp or equivalently over the remnant masses, the compact remnant mass distribution ${\rm d} p/{\rm d} m_\bullet$ appearing in Eq.~(\ref{eq|easybirthrate}) yields only a normalization factor independent of metallicity and SFR. Then the integral involving the SFR becomes simply $\rho_{\psi}$, as in the cosmic case of Eq.~(\ref{eq|cosmicbirthrate}). Some differences between the two approaches show up in the merging rate  $R_{\rm merge}(\mathcal{M}_{\bullet\bullet}, t)$ as a function of the chirp mass (see bottom panels in Fig.~\ref{fig|Rmerg}), since this is determined by the birthrate shape (see Fig.~\ref{fig|Rbirth}).
Specifically, $R_{\rm merge}(\mathcal{M}_{\bullet\bullet}, t)$ from our approach turns out to be shifted toward smaller chirp masses with respect to the cosmic case (due to the higher metallicity occurring in galaxies). We stress that such differences, as well as those in the birthrates, are currently not significant given that they can be hidden behind the overall astrophysical uncertainties.

In Fig.~\ref{fig|Rmerg_3D} we illustrate the probability of merging for compact binaries at $z\sim 0$, $3$ and $6$, as a function of the chirp mass $\mathcal{M}_{\bullet\bullet}$ and of the SFR $\psi$ in the host galaxy progenitor (plainly, if occurring in spheroids the SFR can be much lower at the time of the merger event); the colored surfaces refer to BH-BH, NS-NS and BH-NS events. The dependence in the shape of the surfaces on redshift reflects both the evolution in the SFR functions and the behavior of the Spera et al. (2017) relation $m_\bullet(m_\star|Z)$ at different metallicities. It is worth noticing that for larger chirp masses the local BH-BH merging probability distribution is more extended toward disk galaxy hosts with smaller SFR (see also bottom left panel in Fig.~\ref{fig|Rmerg}), which have also lower metallicity and hence an increased occurrence of massive BH remnants.

\section{GW detection rates from merging binaries in galaxies}\label{sec|GWdetection}

We now turn to compute and discuss the GW detection rates from merging binaries. Taking up the formalism by Taylor \& Gair (2012; see references therein), we compute the GW event detection rate per unit redshift, chirp mass $\mathcal{M}_{\bullet\bullet}$, and signal-to-noise ratio (SNR) $\rho$ as:
\begin{equation}
{{\rm d}\dot{N}\over {\rm d}\mathcal{M}_{\bullet\bullet}\,{\rm d}\rho\,{\rm d}z}(\mathcal{M}_{\bullet\bullet}|\rho,z)={{\rm d} V\over {\rm d}z}\,{R_{\rm merge}(\mathcal{M}_{\bullet\bullet}, z)\over (1+z)}\, P_\rho(\rho|\mathcal{M}_{\bullet\bullet},z)~;
\label{eq|detectiorate}
\end{equation}
here $R_{\rm merge}(\mathcal{M}_{\bullet\bullet}, z)$ is the merging rate per unit chirp mass from Eq.~(\ref{eq|mergerate}), ${\rm d}V/{\rm d}z$ is the comoving volume per unit redshift interval, the factor $1/(1+z)$ takes into account cosmological time dilation, and $P_\rho(\rho| \mathcal{M}_{\bullet\bullet},z)$ is the distribution of SNR at given chirp mass and redshift. The latter quantity is in turn computed as
\begin{equation}
P_\rho(\rho|\mathcal{M}_{\bullet\bullet},z)=P_\Theta(\Theta_\rho)\,{\Theta_\rho\over \rho}~,
\label{eq|Prho}
\end{equation}
in terms of the orientation function
\begin{equation}
\Theta_\rho={\rho\over 8}\,{D_L(z)\over R_0}\,\left[{1.2\,M_\odot\over (1+z)\,\mathcal{M}_{\bullet\bullet}}\right]^{5/6}\,{1\over \sqrt{\zeta_{\rm isco}+\zeta_{\rm insp}+\zeta_{\rm merg}+\zeta_{\rm ring}}}~
\label{eq|thetarho}
\end{equation}
and of its distribution function
\begin{equation}
P_\Theta(\Theta)=\left\{
\begin{aligned}
&5\,\Theta\,(4-\Theta)^3/256  &0<\Theta<4 \\
&0  &{\rm otherwise}~.
\end{aligned}
\right.
\label{eq|ptheta}
\end{equation}
In the above expressions $D_L(z)$ is the luminosity distance from the GW source at redshift $z$. In addition, $R_0$ is the detector characteristic distance parameter given by\footnote{Hereafter $1\, M_\odot\approx 2\times 10^{33}$ g.}
\begin{equation}
R_0^2 = {5\, M_\odot^2\over 192\, \pi\, c^3}\,\left({3\, G\over 20}\right)^{5/3}\, x_{7/3} 
\end{equation}
in terms of the auxiliary quantity
\begin{equation}
x_{7/3} = \int_0^\infty{{\rm d}f\,\over (\pi\, M_\odot)^{1/3}\, f^{7/3}\, S(f)}
\end{equation}
with $S(f)$ the noise power spectral density. Finally,
\begin{equation}
\begin{aligned}
\zeta_{\rm isco} &= {1\over (\pi\, M_\odot)^{1/3}\, x_{7/3}}\, \int_{0}^{2\, f_{\rm isco}}{{\rm d}f\over S(f)}\, {1\over f^{7/3}}\\
\\
\zeta_{\rm insp} &= {1\over (\pi\, M_\odot)^{1/3}\, x_{7/3}}\,\int_{2\, f_{\rm isco}}^{f_{\rm merg}}{{\rm d}f\over S(f)}\, {1\over f^{7/3}}\\
\\
\zeta_{\rm merg} &= {1\over (\pi\, M_\odot)^{1/3}\, x_{7/3}}\,\int_{f_{\rm merg}}^{f_{\rm ring}}{{\rm d}f\over S(f)}\, {1\over f^{4/3}\, f_{\rm merg}}\\
\\
\zeta_{\rm ring} &= {1\over (\pi\, M_\odot)^{1/3}\, x_{7/3}}\,\int_{f_{\rm ring}}^{f_{\rm cut}}{{\rm d}f\,\over S(f)}\, {1\over f_{\rm{merg}}\,f_{\rm ring}^{4/3}}\, \left[1+\left({f-f_{\rm ring}\over \sigma/2}\right)^2\right]^{-2}\\
\end{aligned}\label{eq|zmax}
\end{equation}
are functions specifying the overlap of the waveform with the observational bandwidth during the inspiral ($\zeta_{\rm isco}+\zeta_{\rm insp}$), merger ($\zeta_{\rm merg}$), and ringdown ($\zeta_{\rm ring}$) phases of the event; the above expressions include the phenomenological waveforms by Ajith et al. (2008). In particular, $\zeta_{\rm isco}$ depends on the redshifted frequency at the innermost circular stable orbit $f_{\rm isco}$, which is also
the maximum frequency at which the quadrupolar formula holds; this is given by
\begin{equation}
f_{\rm isco}\simeq {2198\over 1+z}\, \left({M_{\rm bin}\over M_\odot}\right)^{-1}\,\,{\rm Hz}
\end{equation}
where $M_{\rm bin}=\mathcal{M}_{\bullet\bullet}\,(1+q)^{6/5}/q^{3/5}$ is the total mass of the binary (see Finn 1996; Taylor \& Gair 2012). The other parameters $f_{\rm merg}$, $f_{\rm ring}$, $f_{\rm cut}$ and $\sigma$ appearing in Eqs.~(\ref{eq|zmax}) also scale like $M_{\rm bin}^{-1}$, with coefficients weakly depending on the symmetric mass ratio $\eta=q/(1+q)^2$ and possibly on spin, as approximated by Ajith et al. (2008, 2011, 2014).

The GW event detection rates per unit redshift are then obtained by integrating Eq.~(\ref{eq|detectiorate}) over the chirp mass $\mathcal{M}_{\bullet\bullet}$ and SNR $\rho$ from a minimum detection threshold $\rho_0$:
\begin{equation}
{{\rm d}\dot{N}\over {\rm d}z}(>\rho_0,z)=\int_{\rho_0}^\infty {\rm d}\rho\,{{\rm d}\dot{N}\over {\rm d}\rho\,{\rm d}z}(\rho,z)=\int_{\rho_0}^\infty {\rm d}\rho\,\int {\rm d}\mathcal{M}_{\bullet\bullet}\,{{\rm d}\dot{N}\over {\rm d}\mathcal{M}_{\bullet\bullet}\,{\rm d}\rho\,{\rm d}z}(\mathcal{M}_{\bullet\bullet}|\rho,z)~.
\label{eq|zdist}
\end{equation}
Finally, the GW number count rate $\dot N(>\rho_0)$ can be obtained by integrating the above expression over redshift. In Fig.~\ref{fig|GWzdist} we report our results concerning ${\rm d}\dot{N}(>\rho_0)/{\rm d}z$ for AdvLIGO/Virgo (top panel) and for the ET (bottom panel), with minimum SNR $\rho_0=8$. Black solid lines refer to BH-BH events, dashed to NS-NS and dotted to BH-NS ones.

Although the intrinsic merging rate is larger for NS-NS than for BH-BH (cf. Sect.~\ref{sec|mergerrates}) the detector response makes the rate of GW events from BH-BH binaries to overcome that from NS-NS binaries toward increasing redshift; the crossover occurs at $z\sim 0.05$ for AdvLIGO/Virgo and around $z\sim 0.5$ for ET. The increasing dependence of detectability on the chirp mass implies that: GW event rate from BH-BH mergers peaks at $z\approx 0.3-0.4$ and then falls off rapidly at $z\ga 1$ for AdvLIGO/Virgo, while it has a broad shape peaking around $z\approx 1.5$ with an extended tail out to very high redshift for ET; GW event rate from NS-NS mergers can be practically detected only within a few hundred Mpcs for AdvLIGO/Virgo while out to $z\la 2.5$ with ET; GW event rates from BH-NS mergers peak at $z\approx 0.3$ and then steeply fall off for AdvLIGO/Virgo, while they have a more extended redshift distribution for ET, mirroring the shape of the BH-BH rate with a lower normalization.

In the same Fig.~\ref{fig|GWzdist} we compare the GW event rate computed from the star formation and chemical enrichment history of individual galaxies (black lines) vs. the approach based on the cosmic SFR density and cosmic metallicity (green lines). Differences are appreciable in BH-BH and BH-NS rates toward increasing redshift for AdvLIGO/Virgo; e.g., the BH-BH event rate in the cosmic approach relative to our is larger by a factor of $\sim 2$ when integrated over all redshift, by a factor of $\sim 4.5$ when integrated over $z\ga 1$, and by a factor $\sim 15$ when integrated over $z\ga 2$.
These outcomes can be traced back to the dependence of the quantity $R_{\rm merge}(\mathcal{M}_{\bullet\bullet},t)$ entering Eq.~(\ref{eq|detectiorate}) on the chirp mass. Galaxy metallicities are typically larger than the cosmic value, so lowering the formation efficiency of large BH masses and reducing the detectability (given the SNR threshold $\rho_0=8$) of GWs from BH-BH and BH-NS mergers toward high redshift; plainly the rate of light binaries like NS-NS are not affected. For the more sensitive ET the differences in the detected events between the two approaches is negligible at SNR threshold of $8$, since most of the merger events are detected out to high redshifts, though the distribution in chirp masses stays somewhat different (see Fig.~\ref{fig|Rmerg}, bottom panels).

We stress that the performances of current GW instruments allow to probe NS-NS mergers only at very low redshift $z\la 0.1$. The unique event GW170817 detected so far is located at $z\approx 0.01$ (see Abbott et al. 2017b,c); interestingly, its host galaxy NGC4993 is known to be an early-type with no ongoing star formation and old stellar populations with loosely constrained age $\ga 3-6-10$ Gyr (see Im et al. 2017; Troja et al. 2017; Blanchard et al. 2017). It has been pointed out (e.g., Palmese et al. 2017; Belczynski et al. 2018) that finding the very first NS-NS merger within a galaxy with old stellar populations and low SFR may be in tension with theoretical estimates. To check what happens in our framework, we first note that for NS-NS mergers (but not for BH-BH or BH-NS) in Eq.~(\ref{eq|easybirthrate}) the dependence on galaxy metallicity of the remnant mass distribution can be safely neglected (cf. Fig.~\ref{fig|remnant}), so that to a very good approximation $R_{\rm birth, NS-NS}(m_\bullet,t)\simeq \rho_{\psi}(t)\, F(m_\bullet)$ where $F(m_\bullet)$ is solely function of the primary mass; when inserting this expression in Eq.~(\ref{eq|mergratetemp}) and integrating over the relevant range of NS primary masses to find the number of mergers, such function enters just in an overall multiplicative factor $\int{\rm d}m_\bullet\, F(m_\bullet)$.

Thus the fraction of NS-NS mergers occurring at the present cosmic time $t_0$ from galaxies with age older than $T$ is given by
\begin{equation}
f_{\rm NS-NS}(t_0|{\rm age}>T)\simeq \frac{\int_{T}^{t_0}\, {\rm d} t_d\, \rho_\psi(t_0-t_d)\, {{\rm d} p/ {\rm d} t_d}}{\int_{t_{d,{\rm min}}}^{t_0} {\rm d} t_d\, \rho_\psi(t_0-t_d)\, {{\rm d} p/ {\rm d} t_d}}~;
\end{equation}
in the same vein, the fraction of mergers occurring in disks and spheroids can be computed via the same expression, by replacing $\rho_\psi$ at the numerator with the corresponding contribution from these galaxy types (cf. Sect.~\ref{sec|SFR_func} and Fig.~\ref{fig|SFRcosm}). Using the detailed redshift dependent shape of $\rho_\psi(t)$, we have computed that the overall contribution (in discs plus spheroids) to the local NS-NS rate from compact binaries in galaxies older than $3-6-10$ Gyr amounts to $60-45-20\%$, and in particular the contribution from spheroids older than $3-6-10$ Gyr amounts to $52-41-20\%$. For an instructive back-of-the-envelope calculation, one can approximately use in the integrands above the average values  $\langle\rho_\psi\rangle$ over the relevant cosmic time intervals, to obtain $f_{\rm NS-NS}(t_0|{\rm age}>T)\simeq \langle \rho_\psi(t)\rangle_{|t<t_0-T} \ln{(t_0/T)}/\langle\rho_\psi(t)\rangle_{|t<t_0-t_{\rm d,min}}\,\ln{(t_0/t_{\rm d,min})}$; e.g., using $\langle\rho_\psi(t)\rangle_{|t<t_0-T}\approx 0.089-0.12\,M_\odot$ yr$^{-1}$ Mpc$^{-3}$ for $T=3-6-10$ Gyr and $\langle\rho_\psi(t)\rangle_{|t<t_0-t_{\rm d,min}}\approx 0.038\,M_\odot$ yr$^{-1}$ Mpc$^{-3}$, this approximation is seen to produce the same fractions $\approx 60-45-20\%$ of the computation above. Interestingly, these estimated fractions are independent of the parameter $f_{\rm eff}$ entering Eq.~(\ref{eq|easybirthrate}), that as discussed in Sect.~\ref{sec|mergerrates} is challenging to compute ab initio or to constrain on an observational basis. All in all, we expect that a substantial number of NS-NS binaries merging at $z\approx 0$ have been created in the starforming progenitors of local spheroids at appreciably earlier cosmic times. Catching in real time the mergers with a short delay time at redshift $z\approx 2.5$ (where the cosmic SFR density peaks) will likely become achievable with the ET.

In Fig.~\ref{fig|GWzdist_complot} we show how the GW event rate as a function of redshift for AdvLIGO/Virgo depends on some relevant parameters and assumptions used in our computations. In the top left panel, the minimum SNR for detection is varied from our fiducial value $\rho_0=8$ to $5$, to $13$, and to $24$. Plainly, considering larger SNR $\rho_0$ decreases the detectability of the GWs toward higher redshift. In the top middle panel we show the contribution to the event rates from different phases of the compact binary mergers. Our basic computation includes all phases, i.e., the inspiral, the merger, and the ringdown. Removing the ringdown ($\zeta_{\rm ring}$=0) and the merger ($\zeta_{\rm merg}=\zeta_{\rm ring}=0$) plainly keeps only events with inspiral phase crossing the detector bandwidth, so reducing the number of observable high-mass merging binaries and hence their event rates; the outcome is still interesting because for such events the parameter reconstruction from the waveform is expected to be most effective. However, it is worth noticing that in the literature two approximations in estimating event rates are often adopted. The first consists in taking  frequencies up to $f_{\rm isco}$ at which the quadrupolar formula holds, corresponding to keep only $\zeta_{\rm isco}$ (i.e., setting $\zeta_{\rm insp}=\zeta_{\rm merg}=\zeta_{\rm ring}=0$) in Eq.~(\ref{eq|thetarho}); the outcome is found to constitute a conservative lower limit to the event rates. The second consists in setting $\zeta_{\rm insp}=\zeta_{\rm merg}=\zeta_{\rm ring}=0$ and $\zeta_{\rm isco}\simeq 1$, corresponding to consider that the inspiral phase of any event completely overlaps with the detector bandwith; we warn that this approximation actually holds only for NS-NS (see Taylor \& Gair 2012), but for BH-BH and BH-NS it considerably overpredicts the rates, especially toward high-redshift.

In the bottom left panel we vary the IMF from the fiducial shape by Chabrier (2003), to that by Salpeter (1955), by Kroupa (2002), to the top-heavy IMF by Lacey et al. (2010), to the metallicity-dependent IMF by Martin-Navarro et al. (2015). In the bottom middle panel, we show the effect of
including or excluding (P)PSNe from the remnant mass spectrum by Spera et al. (2017). 
In the bottom right panel, we vary the time delay distribution from the fiducial shape ${\rm d}p/{\rm d}t_d\propto t_d^{-1}$ to a somewhat flatter one $\propto t_d^{-0.75}$ and a somewhat steeper one $\propto t_d^{-1.5}$. In all these cases, the shape of the GW event rate $z-$distribution is moderately affected, within factors a few at most; notice that the different curves plotted here have been computed self-consistently by normalizing the corresponding local BH-BH merging rate to the value observed by LIGO/Virgo; thus the reader should keep in mind that they underlie different values of $f_{\rm eff}$.

The (redshift-integrated) Euclidean-normalized GW counts are shown in Fig.~\ref{fig|GWcounts}, both for AdvLIGO/Virgo and ET. Here we just notice that for electromagnetic signals the counts of a uniform distribution of sources with a smooth distribution of luminosities (Euclidean counts) obeys the scaling $N(>S)\propto S^{-3/2}$ in terms of the flux $S$ (e.g., Weinberg 2008); this is basically because $N(>S)\propto V\propto D_L^3$ and $S\propto D_L^{-2}$ hold. In the case of GWs, the relation between SNR and distance is inverse linear $\rho\propto D_L^{-1}$ implying the Euclidean behavior $N(>\rho)\propto \rho^{-3}$ or in differential terms ${\rm d} N/{\rm d}\rho\propto \rho^{-4}$. When this dependence is normalized out, the counts are flat at high SNR which are mainly contributed by local sources, while the decrease toward lower SNRs mainly reflects the rapid evolution in the number density of increasingly distant galaxies.

\subsection{Galaxy-scale gravitational lensing of GW}\label{sec|lensing}

High-redshift $z\ga 2$ star-forming galaxies have a non-negligible probability of being gravitationally lensed by other galaxies (mostly low $z\la 1$ early-types) and by galaxy groups/clusters intervening between the source and the observer (e.g., Blain 1996; Perrotta et al. 2002; Negrello et al. 2007, 2010; Lapi et al. 2012). The GW emission from merging binaries in these sources can be gravitationally lensed too, so enhancing the detectability of high-redshift GW sources (see Ng et al. 2018; Li et al. 2018; Oguri 2018). The effect of a gravitational lensing event with magnification $\mu$ on the GWs emitted by a compact source is to enhance the SNR $\rho\propto\sqrt{\mu}$ without changing the observed frequency structure of the waveform (due to the achromaticity of lensing in the geometrical-optics limit; see Takahashi \& Nakamura 2003). In the following we focus on galaxy-scale gravitational lensing, which is the most efficient for intermediate to high-redshift sources, close to the peak of the cosmic star formation history (see Lapi et al. 2012).

The rate of gravitationally lensed events can be computed as:
\begin{equation}
{{\rm d}\dot{N}_{\rm lensed}\over {\rm d}\rho\, {\rm d} z}=\int^\infty_{\mu_{\rm min}}{\rm d}\mu\,{{\rm d}\dot{N}\over {\rm d}\rho\, {\rm d} z}(\rho/\sqrt{\mu},z)\,{{\rm d} p\over {\rm d}\mu}(\mu,z)~,
\label{eq|lensing}
\end{equation}
where ${\rm d}\dot{N}/{\rm d}\rho\, {\rm d} z$ is the unlensed statistics in Eq.~(\ref{eq|zdist}), and ${\rm d} p/{\rm d}\mu (\mu,z)$ is a probability distribution of amplification factors which depends on the redshift of the GW source and on the properties of the intervening galaxies acting as lenses. The minimum amplification $\mu_{\rm min}$ defines the strength of the lensing events under consideration. We use the amplification distribution derived by Lapi et al. (2012), which takes into account the redshift-dependent statistics of galactic halos, their inner radial distribution of dark matter and baryons, and possible non axisymmetric structure.

The redshift distribution of GW events above a detection threshold $\rho_0$ is
\begin{equation}
{{\rm d}\dot{N}_{\rm lensed}\over {\rm d} z}(>\rho_0)=\int_{\rho_0}^\infty {\rm d}\rho\,{{\rm d}\dot{N}_{\rm lensed}\over {\rm d}\rho\, {\rm d} z}~,
\label{eq|lensingrate}
\end{equation}
and the lensed counts are instead obtained by integrating Eq.~(\ref{eq|lensing}) over redshift. Our results concerning the lensed GW redshift distribution and counts are shown as orange lines in Figs.~\ref{fig|GWzdist} and \ref{fig|GWcounts}; for clarity we illustrate the case $\mu_{\rm min}=10$ to better highlight the overall impact of strong lensing events. Plainly, strongly lensed events have a redshift distribution shifted toward high redshift.

GWs from NS-NS mergers, that in the unlensed case are detectable only locally with AdvLIGO/Virgo and to intermediate redshifts with ET, can in principle be revealed out to $z\la 1$ for AdvLIGO/Virgo and out to high $z$ with ET; however, the lensed rates are very small $\la 10^{-3}$ events per yr with AdvLIGO/Virgo, while they attain even $1$ event per yr with the ET. For AdvLIGO/Virgo lensed GWs rate from BH-BH peak around $z\approx 2$ and attain $\sim 0.1$ event per yr at $z\sim 1-4$, overwhelming the unlensed events for $z\ga 3$; for ET instead the lensed BH-BH rates are of the same order of the lensed NS-NS ones, still factors $\ga 10^3$ below the unlensed. The lensed BH-NS rates feature a similar behavior to the lensed BH-BH, with a lower normalization.

In the top right panel of Fig.~\ref{fig|GWzdist_complot}, we show for AdvLIGO/Virgo how the redshift distribution of gravitationally lensed events is affected  by varying the minimum amplification from our fiducial value $\mu_{\rm min}\approx 10$ to $2$ (defining the strong lensing limit) and to $30$ (a maximal value applying to moderately extended sources, see Lapi et al. 2012); plainly, lowering the minimum amplification yields generally higher lensing rates, though decreasing them toward very high redshift for small chirp mass systems like NS-NS.

We stress that the detection of high-redshift, strongly lensed events can be particularly important for cosmological studies, related to the detection of multiple images and to the characterization of GW time delay distributions (e.g., Lapi et al. 2012; Eales 2015). This is especially true if there is an accompanying electromagnetic emission (e.g., from BH-NS or NS-NS mergers) that can provide independent measurement of the source redshift, and thus help in removing the well-known degeneracy $\rho\propto \sqrt{\mu}\, \mathcal{M}_{\bullet\bullet}^{5/6}/D_L(z)$ among chirp mass, redshift and lensing magnification.

\section{GW background from merging binaries in galaxies}\label{sec|GWback}

The incoherent superposition of weak, undetected GW sources originates a stochastic background (see Abbott et al. 2017f, 2018). In this section we aim to estimate the contribution to such a background by mergers of compact binaries in galaxies. We compute the background energy density at given observed frequency $f_{\rm obs}$ as:
\begin{equation}
\Omega_{GW}(f_{\rm obs})={8\pi\, G\,f_{\rm obs}\over 3\,H_0^3\,c^2}\,\int{\rm d}z\, \int {\rm d}\mathcal{M}_{\bullet\bullet}\,{R_{\rm merge}(\mathcal{M}_{\bullet\bullet}, z)\over (1+z)\, h(z)}\,{{\rm d} E\over {\rm d}f}(f|\mathcal{M}_{\bullet\bullet})\,\int_{\rho<\rho_0}{\rm d}\rho\,P_\rho(\rho|\mathcal{M}_{\bullet\bullet}, z)~,
\label{eq|GWback}
\end{equation}
with $h(z)\equiv [\Omega_M\, (1+z)^3+1-\Omega_M]^{1/2}$. The GW energy spectrum ${\rm d} E/{\rm d}f$ emitted by the binary is taken as (e.g., Zhu et al. 2011)
\begin{equation}
{{\rm d} E\over{\rm d}f}\simeq {(\pi G)^{2/3}\,\mathcal{M}_{\bullet\bullet}^{5/3}\over 3} \times
\left\{
\begin{aligned}
&f^{-1/3} &f<f_{\rm merg}\\
&f_{\rm merg}^{-1}\,f^{2/3} &  f_{\rm merg}\leq f<f_{\rm ring}\\
&{f_{\rm merg}^{-1}\,f_{\rm ring}^{-4/3}\,f^2\over \left[1+\left({f-f_{\rm ring}\over \sigma/2}\right)^2\right]^2} &f_{\rm ring}\leq f<f_{\rm cut}~,
\end{aligned}
\right.
\label{eq|GWenergy}
\end{equation}
in terms of the same parameters $f_{\rm merg}$, $f_{\rm ring}$, $f_{\rm cut}$, and $\sigma$ appearing in Eqs.~(\ref{eq|zmax}).

The results for the stochastic background originated by BH-BH, NS-NS and BH-NS mergers in galaxies are shown in Fig.~\ref{fig|GWback} as black lines, both for AdvLIGO/Virgo and ET.  The thick cyan lines reports the $1\sigma$ sensitivity curves for $1$ yr of observations and co-located detectors (Abbott et al. 2017f; Thrane \& Romano 2013; Crocker et al. 2017). The stochastic background due to BH-BH in galaxies may only marginally be revealed by AdvLIGO/Virgo, while that from all kind of compact binary mergers should be detected with, and possibly characterized by the ET.

\section{Summary}\label{sec|summary}

\begin{figure*}
\centering
\includegraphics[width=16cm]{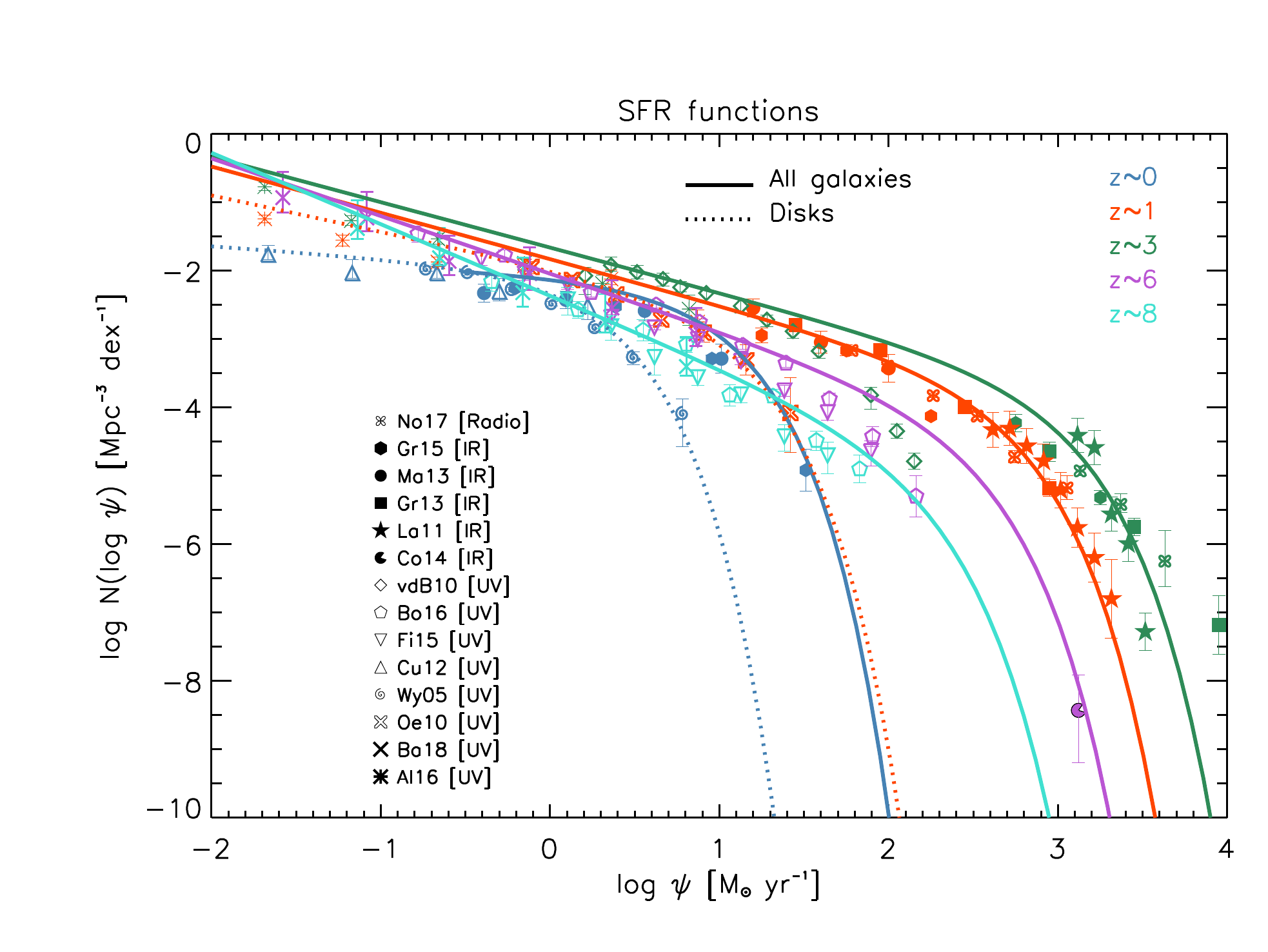}
\caption{The SFR functions at redshifts $z=0$ (blue), $1$ (red), $3$ (green), $6$ (magenta), and $8$ (cyan) by Lapi et al. (2017a,b). Solid lines show the rendition from UV plus far-IR/(sub)mm/radio data, referring to the overall population of galaxies; dotted lines (only plotted at $z\approx 0$ and $1$) show the rendition from (dust-corrected) UV data, referring to disk galaxies. UV data (open symbols) are from van der Burg et al. (2010; diamonds), Bouwens et al. (2016, 2017; pentagons), Finkelstein et al. (2015; inverse triangles), Cucciati et al. (2012; triangles), Wyder et al. (2005; spirals), Oesch et al. (2010; crosses), Alavi et al. (2016; asterisks), Bhatawdekar et al. (2019; X signs); far-IR/(sub)mm data from Gruppioni et al. (2015; hexagons), Magnelli et al. (2013; circles), Gruppioni et al. (2013; squares), Lapi et al. (2011; stars), and Cooray et al. (2014; pacmans); radio data from Novak et al. (2017; clovers).}\label{fig|SFRfunc}
\end{figure*}

\begin{figure*}
\centering
\includegraphics[width=16cm]{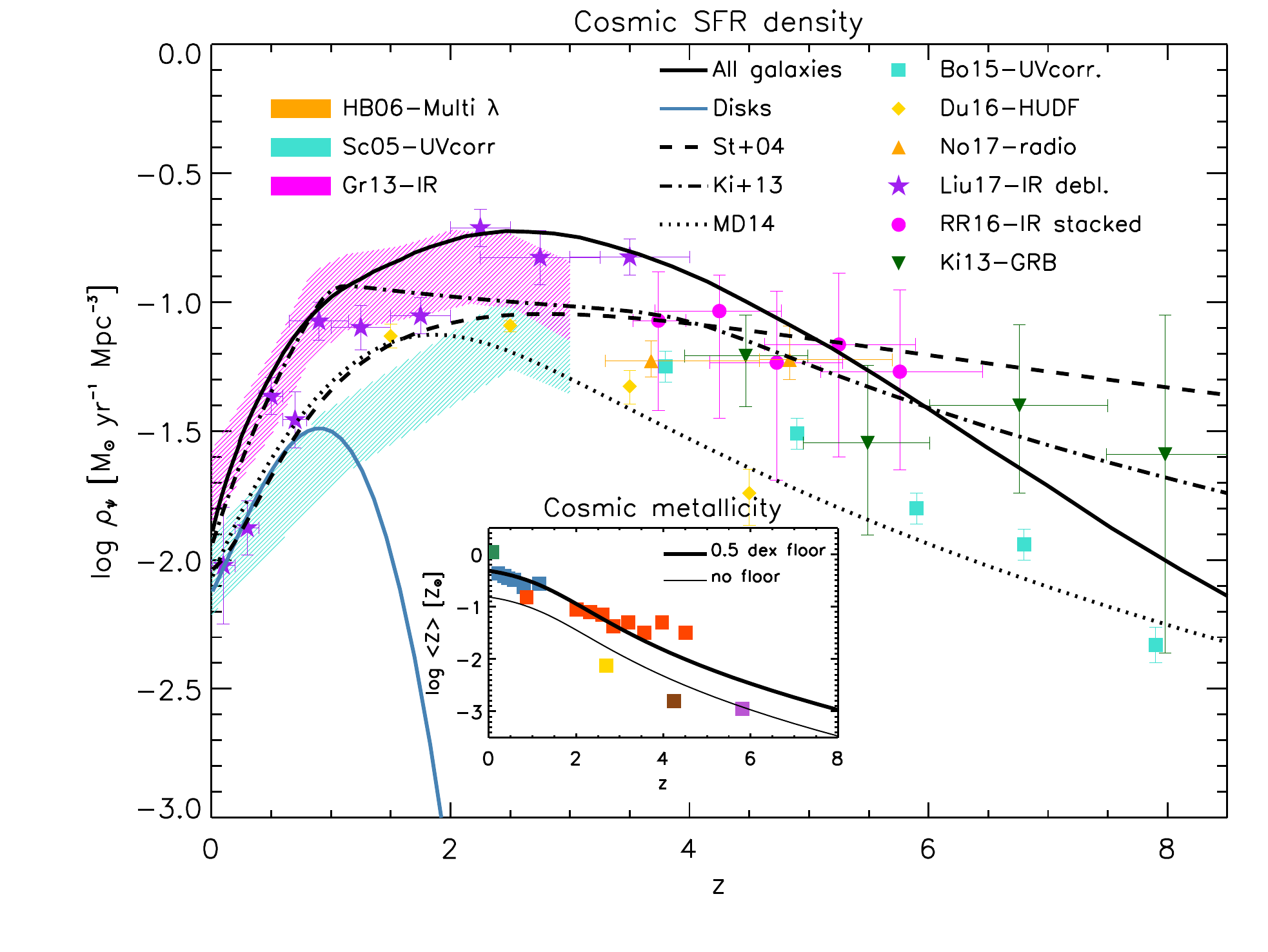}
\caption{Cosmic SFR density as a function of redshift. The black solid line shows the overall results derived from integrating the total UV+far-IR/(sub)mm/radio SFR functions.  The contributions from disk galaxies at $z\la 2$ is highlighted by the blue solid line. For reference, the dashed line illustrates the determination by Strolger et al. (2004), the dot-dashed line that by Kistler et al. (2013), and the dotted line that by Madau \& Dickinson (2014). Data are from (dust-corrected) UV observations by Schiminovich et al. (2005; cyan shaded area) and Bouwens et al. (2015; cyan squares); ALMA submm observations of UV-selected galaxies on the HUDF by Dunlop et al. (2017); VLA radio observations on the COSMOS field by Novak et al. (2017); multiwavelength determination including UV, radio, H$\alpha$, and mid-IR $24\, \mu$m data collected by Hopkins \& Beacom (2006; orange shaded area); Herschel far-IR observations by Gruppioni et al. (2013 magenta shaded area); Herschel far-IR stacking by Rowan-Robinson (2016; magenta circles); far-IR/(sub)mm observations from super-deblended data on the GOODS field by Liu et al. (2018); and estimates from long GRB rates by Kistler et al. (2009; 2013; green stars). The inset shows the cosmic metallicity computed according to Eq.~(\ref{eq|cosmicmetallicity}; thin line), after a floor value of 0.5 dex has been applied (thick line), against the observational constraints collected by Madau \& Dickinson (2014 and references therein; also Aguirre et al. 2008; Balestra et al. 2007; Rafelski et al. 2012).}\label{fig|SFRcosm}
\end{figure*}

\begin{figure*}
\centering
\includegraphics[width=16cm]{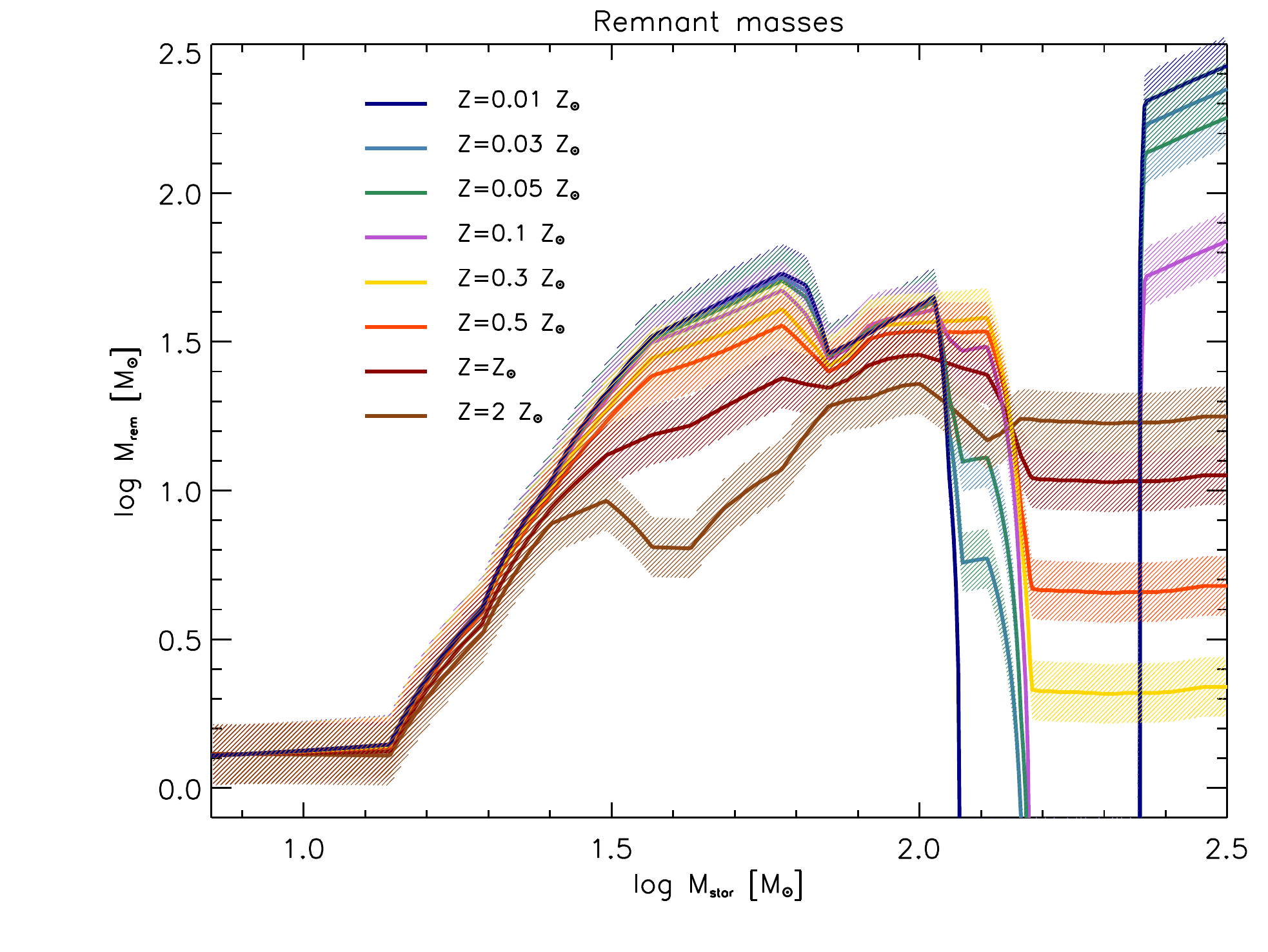}
\caption{Compact remnant mass as a function of the zero-age main sequence star mass at different metallicities $Z=0.01\, Z_\odot$ (blue lines), $0.03\, Z_\odot$ (cyan lines), $0.05\, Z_\odot$ (green lines), $0.1\, Z_\odot$ (magenta lines), $0.3\, Z_\odot$ (yellow lines), $0.5\, Z_\odot$ (orange lines), $Z_\odot$ (brown lines), $2\, Z_\odot$ (saddle brown lines). Solid lines illustrates the relation $m_{\bullet}(m_\star,Z)$ by Spera et al. (2017) for single stellar evolution, based on the delayed SN engine and including (P)PSNe. We have adopted a compact mass remnant distribution ${\rm d}p/{\rm d}\log m_\bullet$ with a log-normal shape centered around this relation and with a $1-\sigma$ dispersion of $0.1$ dex (illustrated by the shaded areas; see Eq.~\ref{eq|remnant}).}\label{fig|remnant}
\end{figure*}

We have investigated the merging rates of compact binaries in galaxies, and the related rates of GW detection events with AdvLIGO/Virgo and with the Einstein Telescope. We have based our analysis on three main ingredients (see Sect.~\ref{sec|basics}): (i) redshift-dependent galaxy statistics provided by the latest determination of the SFR functions from UV+far-IR/(sub)mm/radio data (see Sect.~\ref{sec|SFR_func} and Fig.~\ref{fig|SFRfunc}); (ii) star formation and chemical enrichment histories for individual galaxies, modeled on the basis of observations (see Sect.~\ref{sec|SFR_hist}); (iii) compact remnant mass distribution and prescriptions for merging of compact binaries from stellar evolution simulations (see Sect.~\ref{sec|stellarevo} and Fig.~\ref{fig|remnant}).

\begin{figure*}
\centering
\includegraphics[width=16cm]{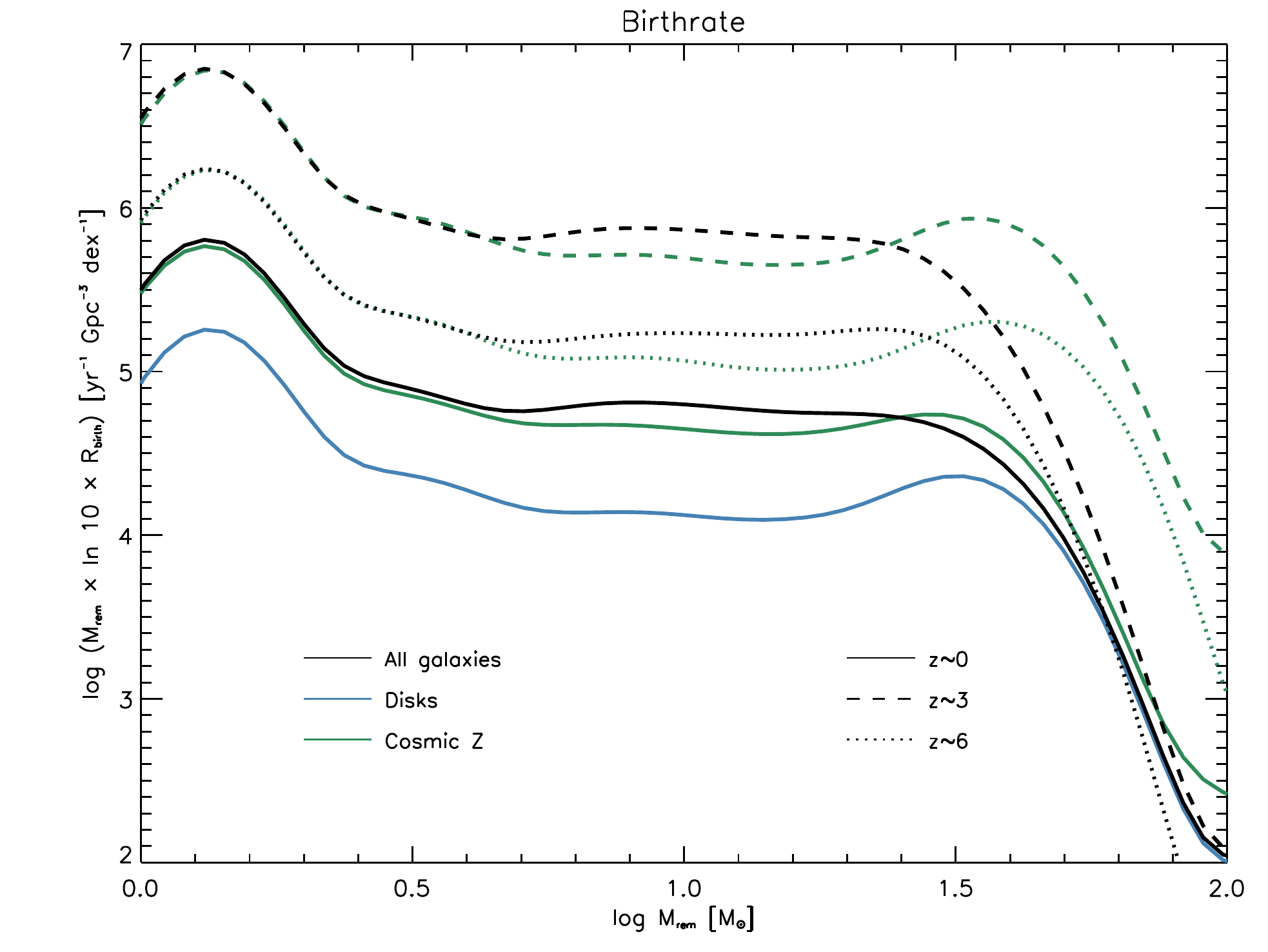}
\caption{Compact remnants birthrate $R_{\rm birth}(\log m_\bullet,z)$ at different redshift $z\sim 0$ (solid lines), $3$ (dashed line), and $6$ (dotted line). Green lines refer to the approach of Eq.~(\ref{eq|cosmicbirthrate}) based on the cosmic SFR density and cosmic metallicity, while black lines refer to our computation of Eq.~(\ref{eq|easybirthrate}) taking into account redshift-dependent galaxy statistics and the star formation and chemical enrichment histories of individual galaxies (blue lines refer to disk-dominated galaxies at $z\la 2$).}\label{fig|Rbirth}
\end{figure*}

\begin{figure*}
\centering
\includegraphics[width=16cm]{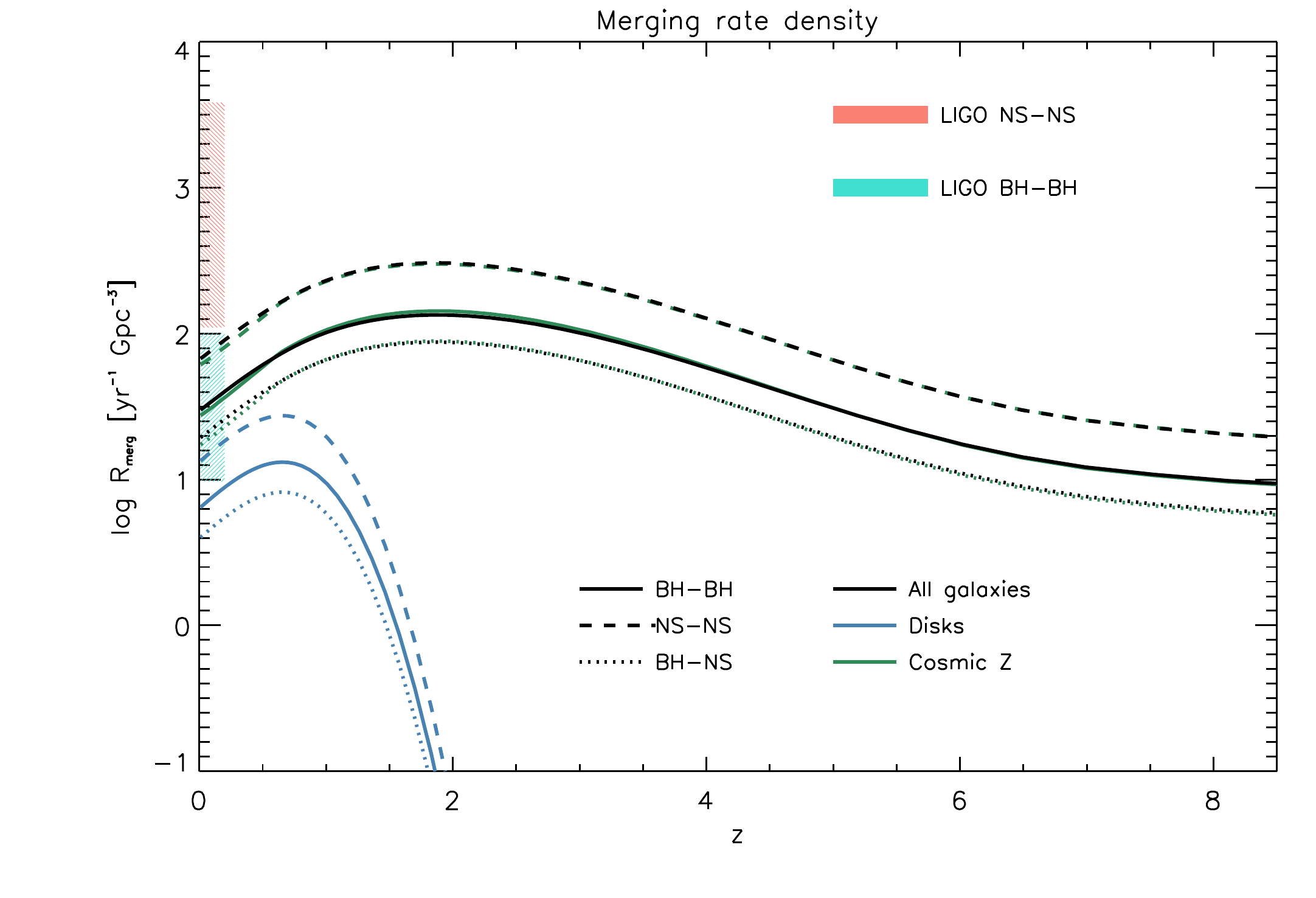}
\includegraphics[width=16cm]{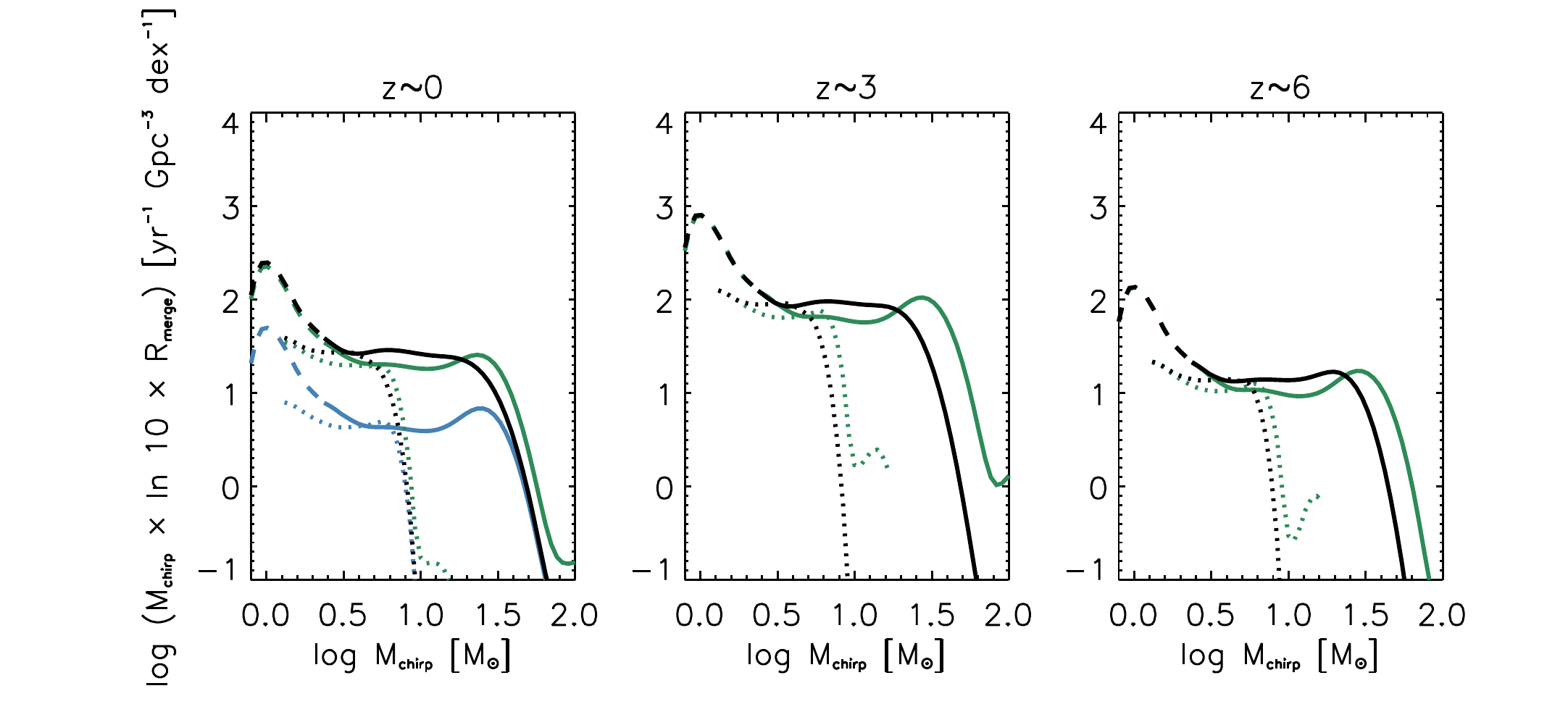}
\caption{Top panel: merging rate density of compact binaries $R_{\rm merg}(z)$ as a function of redshift. Solid lines refer to BH-BH, dashed lines to NS-NS and dotted lines to BH-NS events. Color-code as in Fig.~\ref{fig|Rbirth}. The cyan and orange shaded areas illustrate the local BH-BH and NS-NS merging rates estimated by LIGO in the O2 run, respectively. Bottom panels: merging rate $R_{\rm merge}(\mathcal{M}_{\bullet\bullet},z)$ as a function of the chirp mass at redshift $z\sim 0$ (left), $3$ (middle), and $6$ (right).}\label{fig|Rmerg}
\end{figure*}

\begin{figure*}
\centering
\includegraphics[width=9cm]{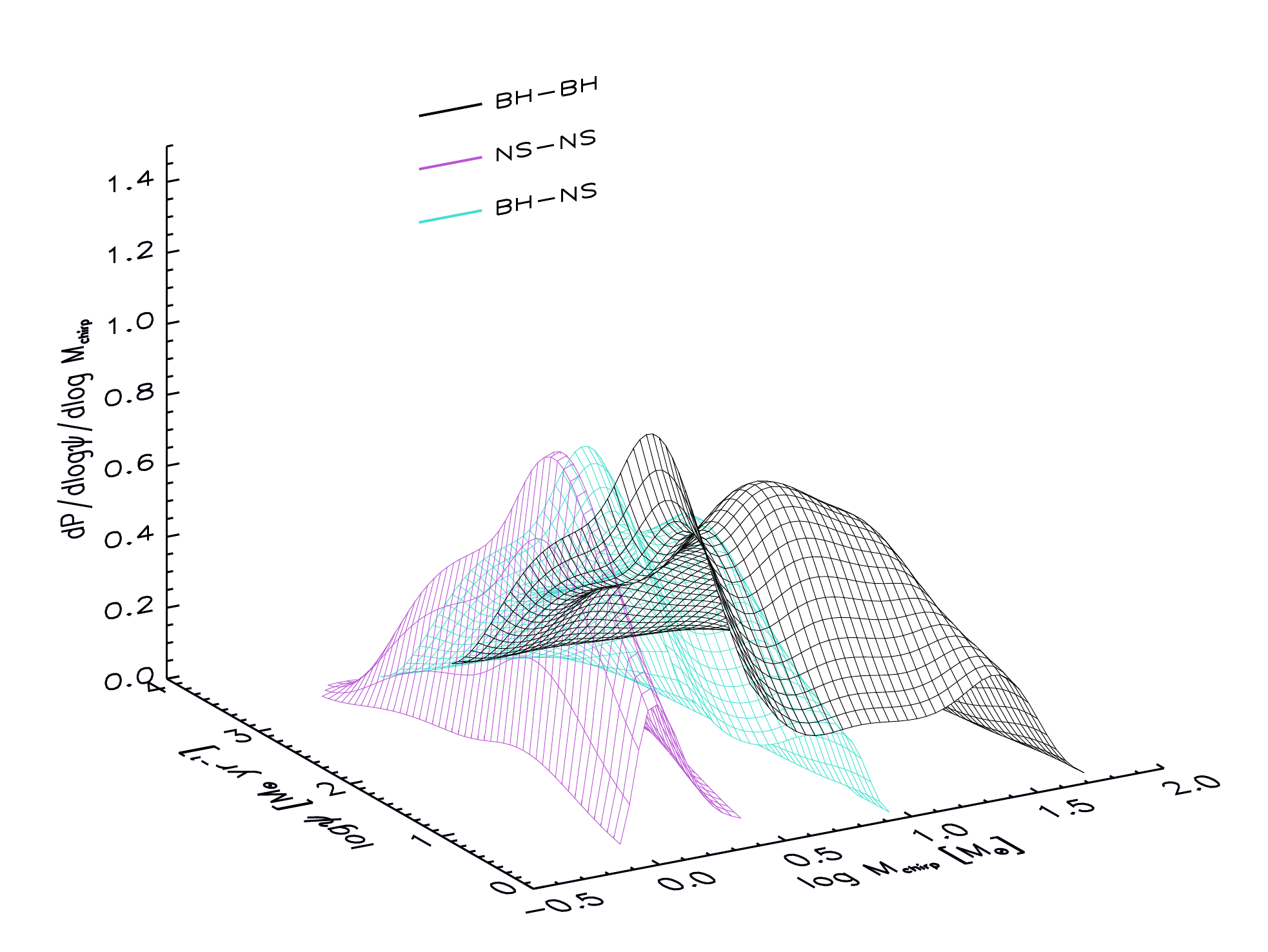}
\includegraphics[width=9cm]{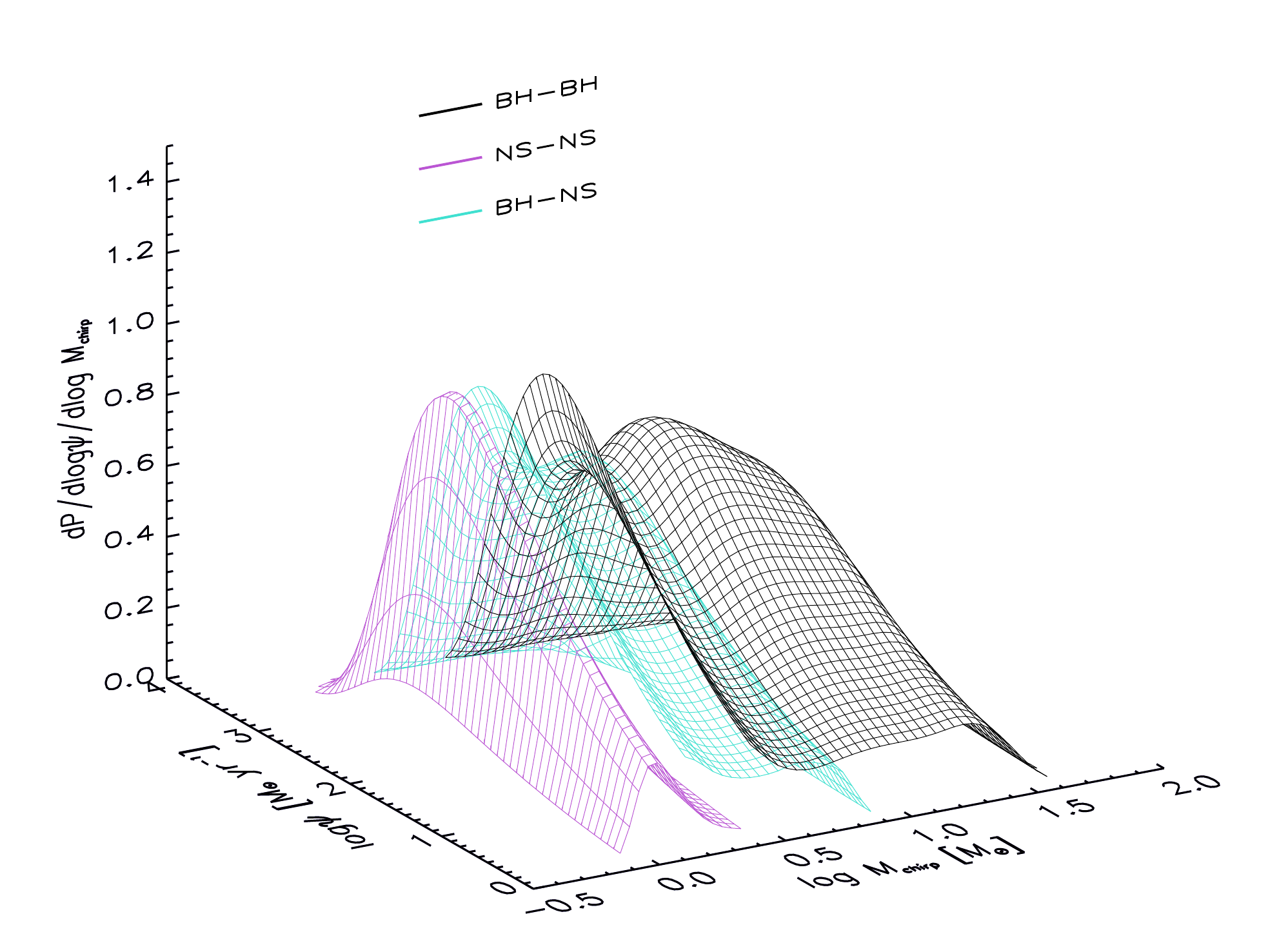}
\includegraphics[width=9cm]{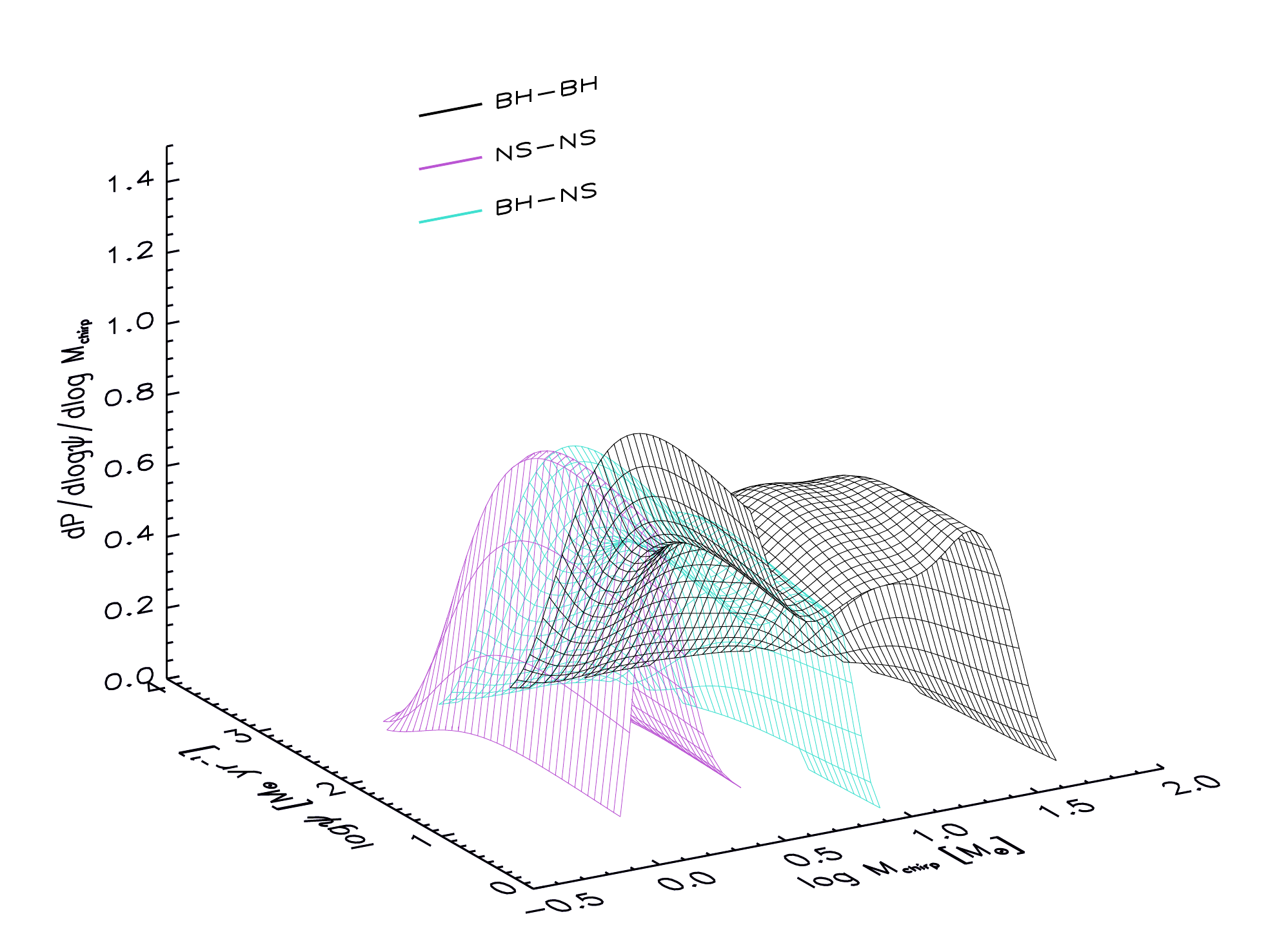}
\caption{Joint probability distribution ${\rm d}P/{\rm d}\log \mathcal{M}_{\bullet\bullet}\,{\rm d}\log\psi$ of chirp masses for merging compact binaries and SFR of the host galaxy progenitor, at redshift
 $z\sim 0$ (top), $3$ (middle), and $6$ (bottom); each surface is normalized to its maximum value. Black surfaces refer to BH-BH, magenta to NS-NS and turquoise to BH-NS events.}\label{fig|Rmerg_3D}
\end{figure*}

\begin{figure*}
\centering
\includegraphics[width=11cm]{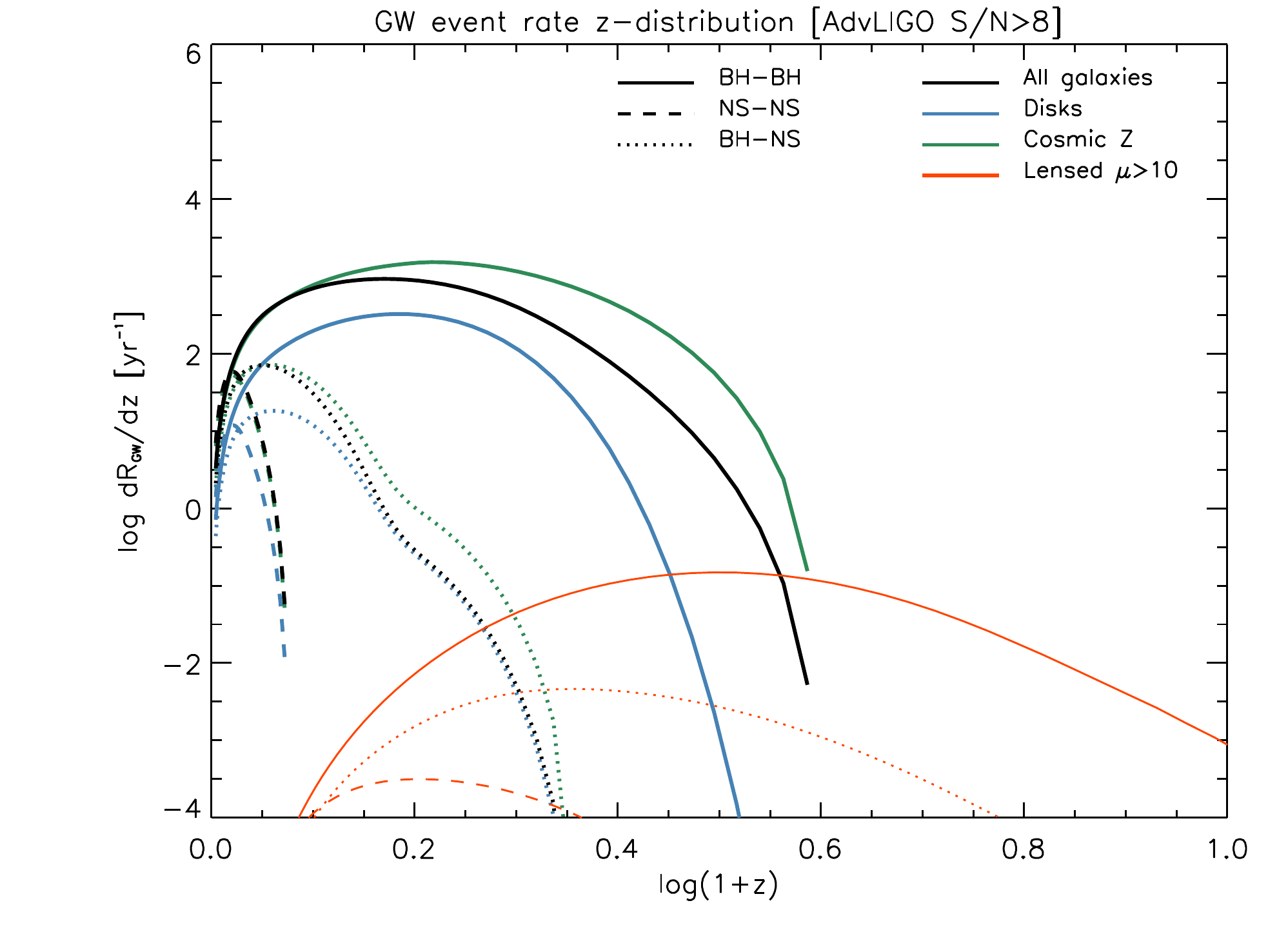}
\includegraphics[width=11cm]{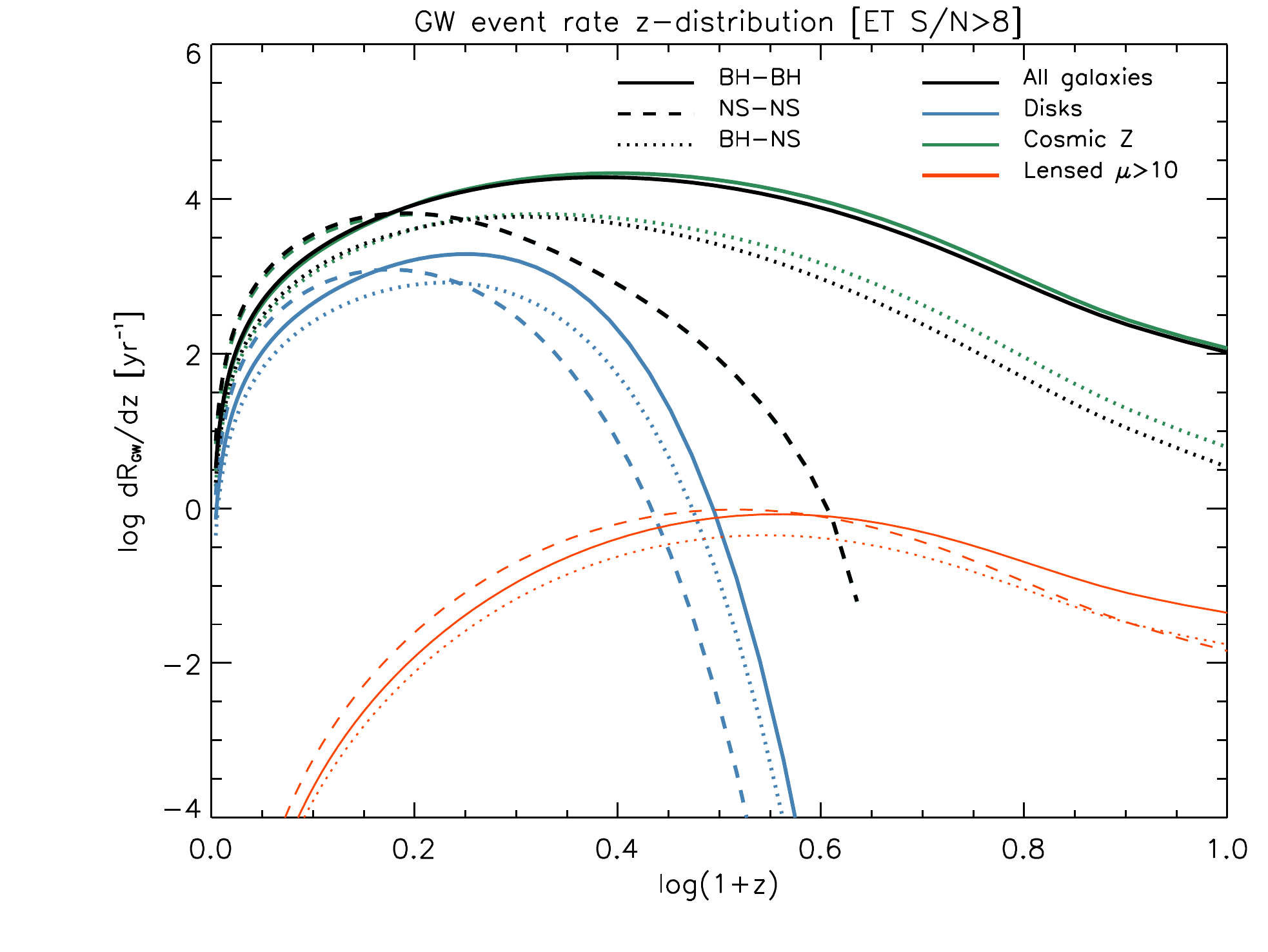}
\caption{GW event rate per unit redshift expected for the AdvLIGO/Virgo (top panel) and ET (bottom panel) with SNR threshold $\rho\ga 8$ (see Sect.~\ref{sec|GWdetection} for details). Linestyles and color-code as in Fig.~\ref{fig|Rbirth}. The orange lines refer to galaxy-scale gravitational lensing of GWs with magnification $\mu\ga 10$ (see Sect.~\ref{sec|lensing}).}\label{fig|GWzdist}
\end{figure*}

\begin{figure*}
\centering
\includegraphics[width=\textwidth]{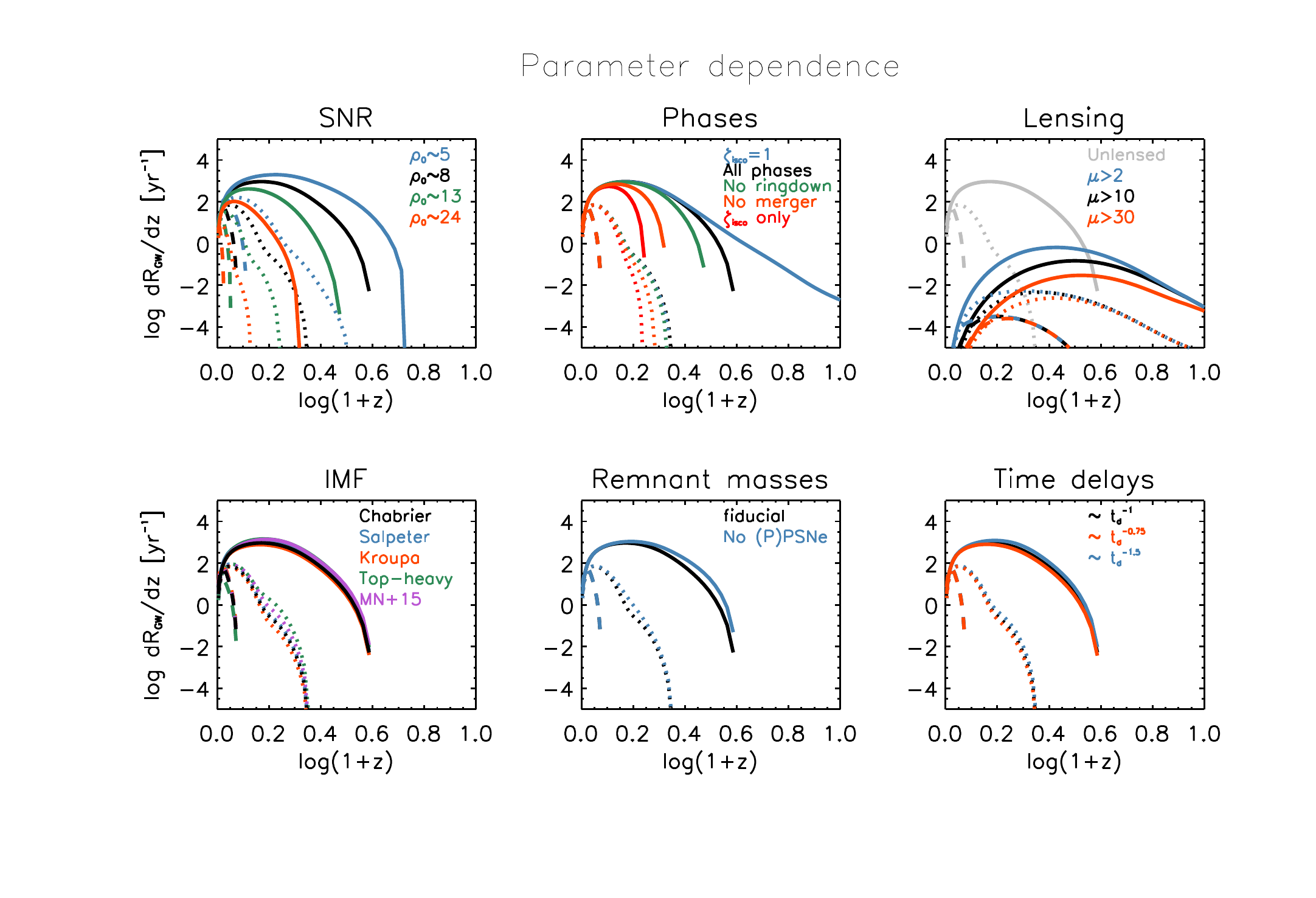}
\caption{Parameter dependence of the GW event rate for AdvLIGO/Virgo. Top left panel: detection SNR $\rho_0$; black lines refer to our fiducial $\rho_0=8$, blue to $\rho_0=5$, green to $\rho_0=13$ and orange to $\rho_0=24$. Top middle right: contribution from different merging phases; black lines include all phases (i.e., inspiral, merger, and ringdown), green lines refer to no ringdown ($\zeta_{\rm ring}=0$ in Eq.~\ref{eq|thetarho}), orange lines to no merger and ringdown ($\zeta_{\rm merg}=\zeta_{\rm ring}=0$ in Eq.~\ref{eq|thetarho}), red lines to keeping inspiral phase up to the ISCO frequency ($\zeta_{\rm insp}=\zeta_{\rm merg}=\zeta_{\rm ring}=0$ in Eq.~\ref{eq|thetarho}), and blue line to assuming complete overlap of the inspiral phase of any event with the detector bandwidth ($\zeta_{\rm insp}=\zeta_{\rm merg}=\zeta_{\rm ring}=0$ and $\zeta_{\rm isco}\simeq 1$ in Eq.~\ref{eq|thetarho}). Top right panel: minimum amplification in lensed counts; solid lines refer to $\mu\ga 10$, blue to $\mu>2$, red to $\mu>30$, while the unlensed statistics are in grey. Bottom left panel: IMF; black lines refer to our fiducial Chabrier (2003), blue to Salpeter (1955), red to Kroupa (2002), green to the top-heavy IMF by Lacey et al. (2010), and magenta to the $Z$-dependent IMF by Martin-Navarro et al. (2015). Bottom middle panel: remnant mass spectrum; black lines refer to the Spera et al. (2017) spectrum when including (P)PSNe, and blue when excluding (P)PSNe. Bottom right panel: delay time distribution; black lines refer to our fiducial distribution ${\rm d}p/{\rm d}t_d)\propto t_d^{-1}$, blue to a steeper one $\propto t_d^{-1.5}$ and orange to a flatter one $\propto t_d^{-0.75}$. In all panels different linestyles show the BH-BH (solid), NS-NS (dashed), and BH-NS (dotted) events.}\label{fig|GWzdist_complot}
\end{figure*}

\begin{figure*}
\centering
\includegraphics[width=11cm]{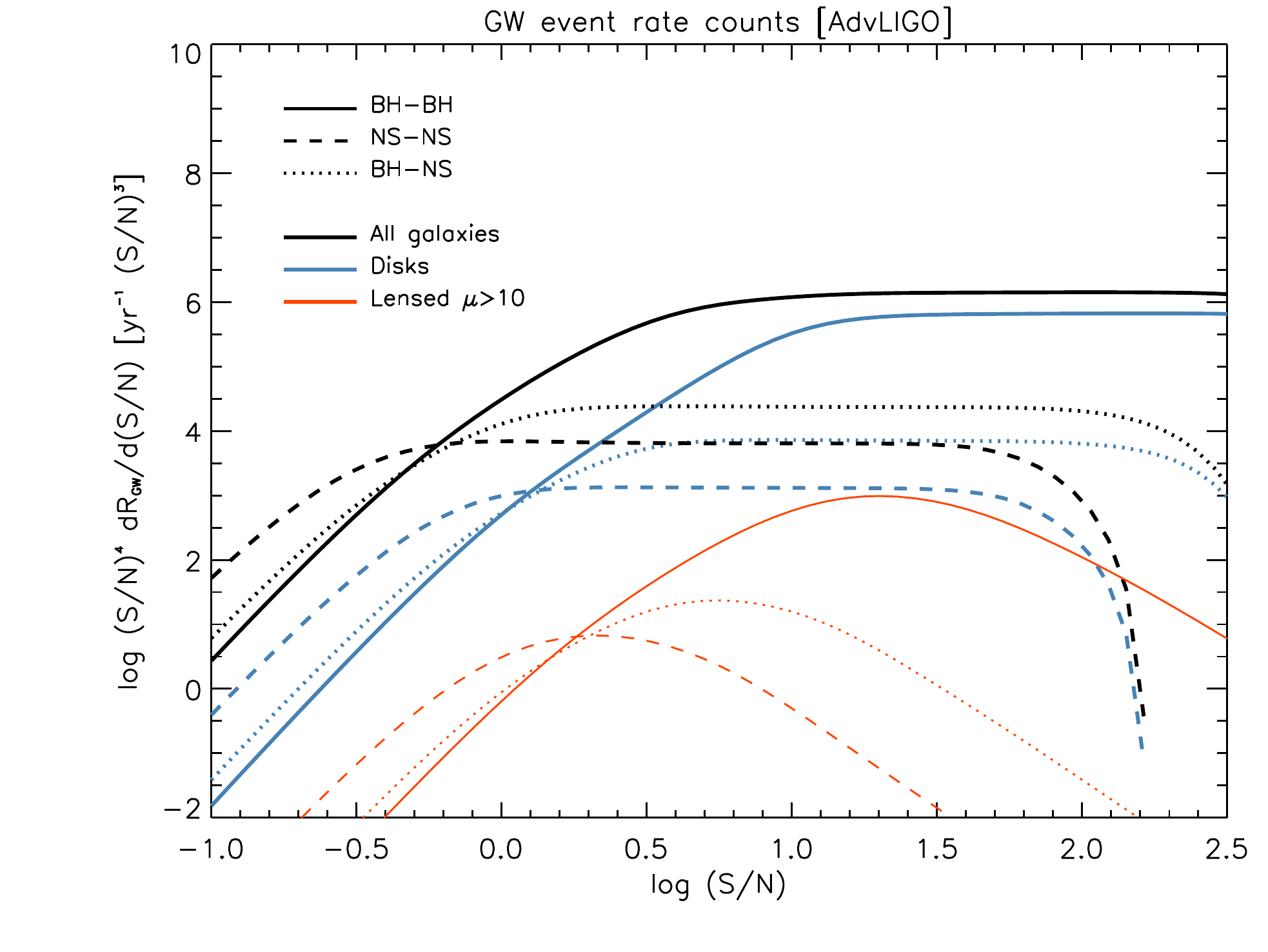}
\includegraphics[width=11cm]{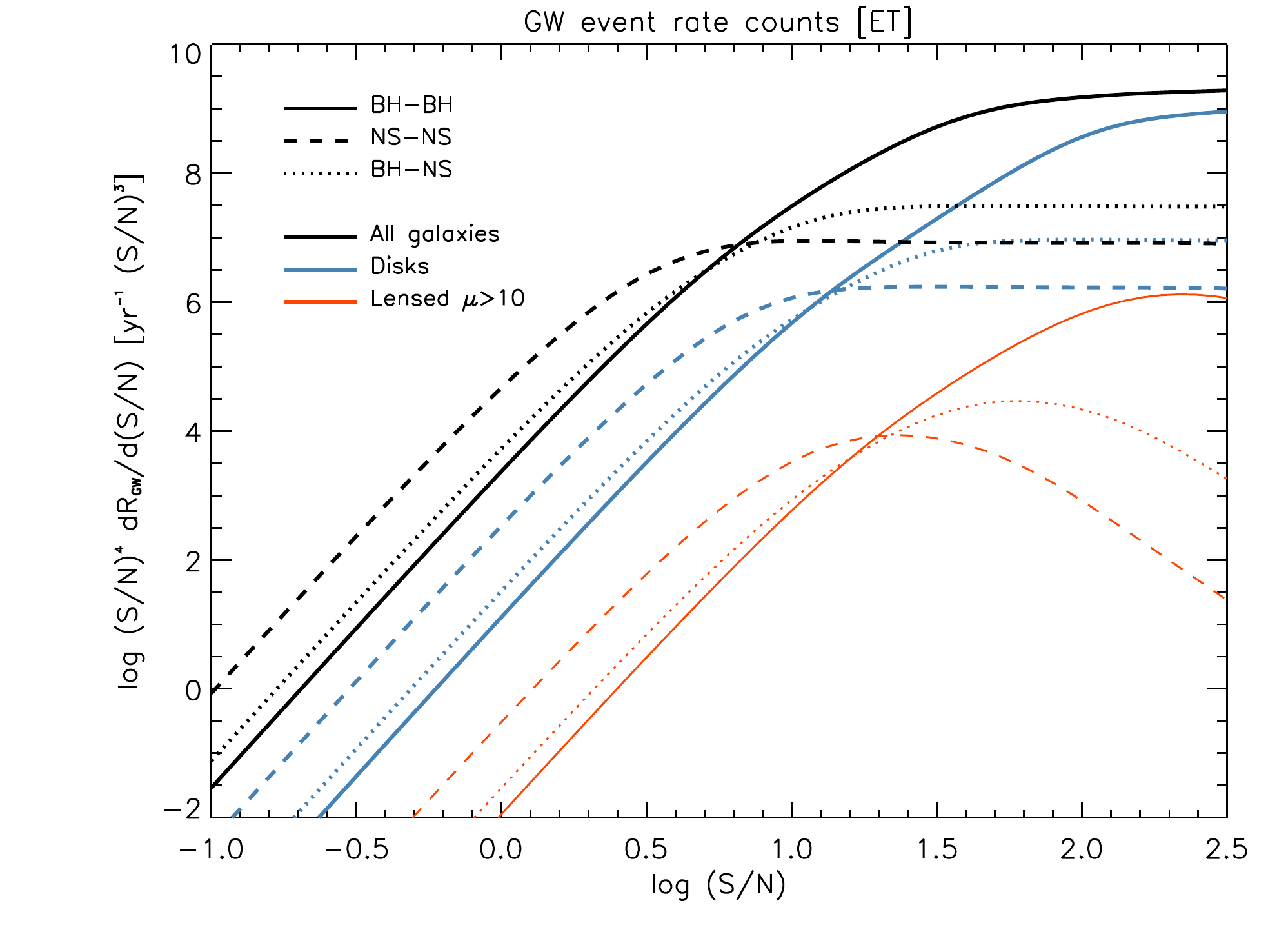}
\caption{Euclidean-normalized counts of GW event rate as a function of the SNR (see Sect.~\ref{sec|GWdetection}) for AdvLIGO/Virgo (top panel) and ET (bottom panel). Linestyles and color-code as in Fig.~\ref{fig|Rbirth}.}\label{fig|GWcounts}
\end{figure*}

\begin{figure*}
\centering
\includegraphics[width=11cm]{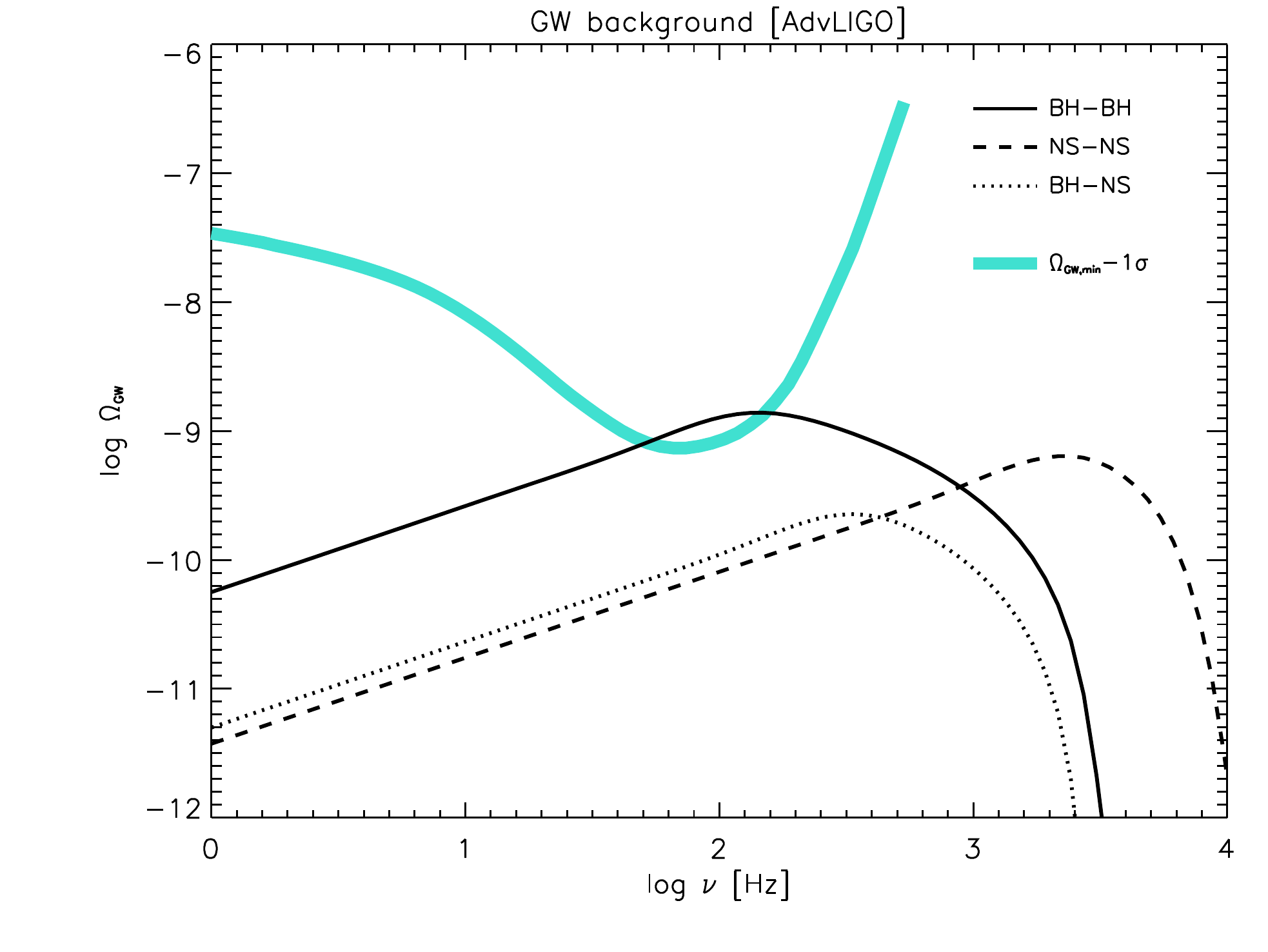}
\includegraphics[width=11cm]{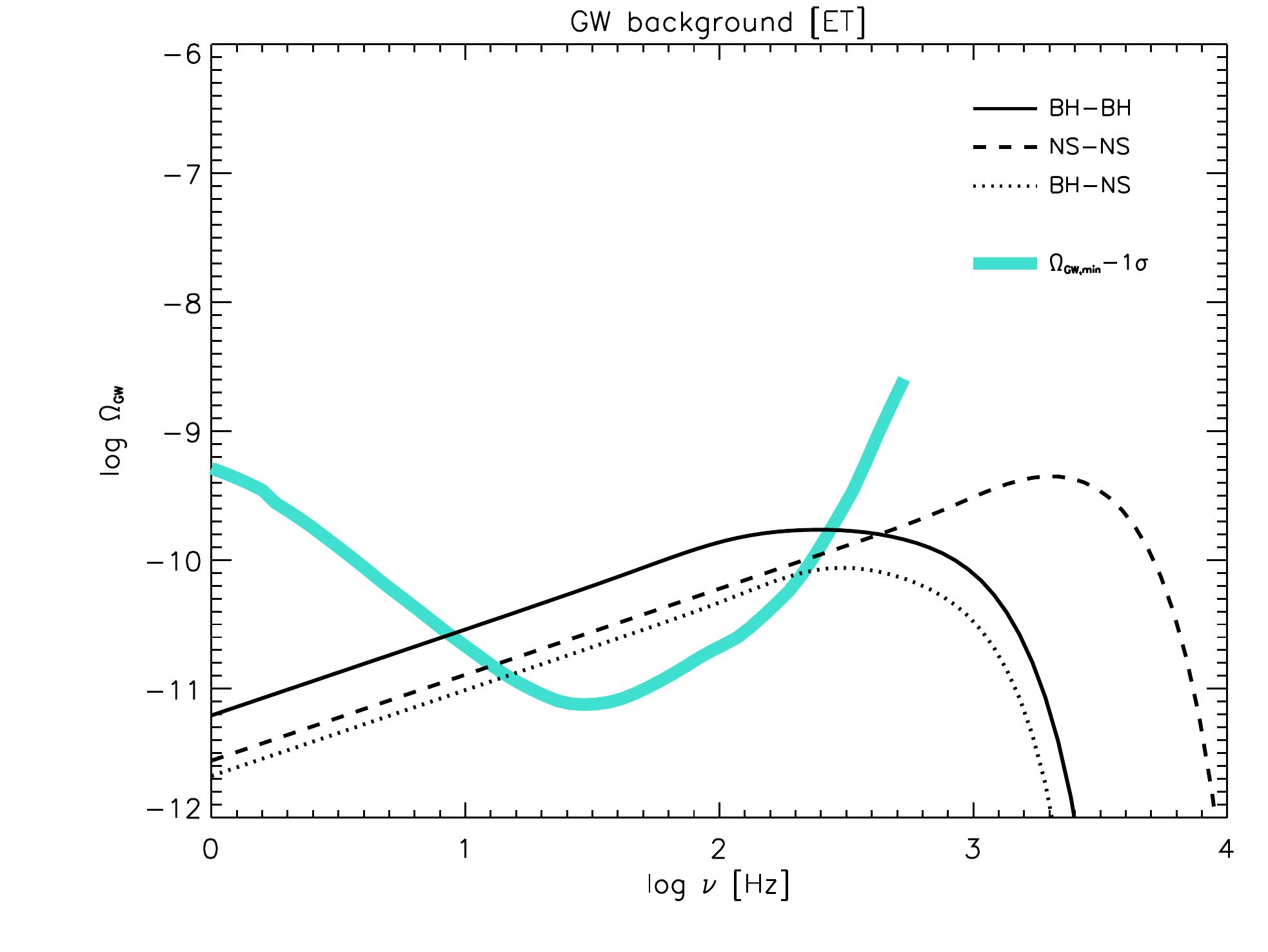}
\caption{Energy density of the GW background as a function of the observed frequency (see Sect.~\ref{sec|GWback}) for AdvLIGO/Virgo (top panel) and ET (bottom panel). Linestyles and color-code as in Fig.~\ref{fig|Rbirth}. The thick cyan lines illustrates $1\sigma$ sensitivity curves for $1$ yr of observations and co-located detectors.}\label{fig|GWback}
\end{figure*}

We have presented results for the intrinsic birthrate of compact remnants (see Sect.~\ref{sec|birthrates} and Fig.~\ref{fig|Rbirth}), the merging rates of compact binaries (see Sect.~\ref{sec|mergerrates} and Fig.~\ref{fig|Rmerg}), and the related GW detection rates and counts (see Sect.~\ref{sec|GWdetection} and Figs.~\ref{fig|GWzdist}, \ref{fig|GWcounts}), attempting to differentiate the outcomes for BH-BH, NS-NS, and BH-NS mergers.
We have compared our approach with the one based on cosmic SFR density and cosmic metallicity, exploited by many literature studies; the merging rates from the two approaches are in agreement within the overall astrophysical uncertainties. We have computed the joint probability distribution of chirp masses related to mergers of compact binaries, and SFR (or stellar mass, metallicity, etc.) of the host galaxy progenitor as a function of redshifts (see Sect.~\ref{sec|mergerrates} and Fig.~\ref{fig|Rmerg_3D}). We have then investigated the impact of galaxy-scale strong gravitational lensing in enhancing the GW event rate of detectable events toward high-redshift (see Sect.~\ref{sec|lensing} and Figs.~\ref{fig|GWzdist}, \ref{fig|GWcounts}). Finally, we have discussed the contribution of undetected GW emission from compact binary mergers to the stochastic background (see Sect.~\ref{sec|GWback} and Fig.~\ref{fig|GWback}).

In a nutshell, our work has been mainly focused on developing an approach to post-process the outcomes of stellar evolution simulations toward computing GW event rates of compact binary mergers (both intrinsic and strongly gravitationally lensed). Specifically, we have coupled the metallicity-dependent compact remnant mass spectrum from stellar evolution simulations to the most recent observational determinations of the galaxy SFR functions and to the star formation and chemical enrichment histories of individual galaxies; such an approach in principle adds extra layers of information with respect to methods based on the integrated cosmic SFR density and cosmic metallicity, like potentially the association of the GW event to the properties of the host galaxy; admittedly, this is a first step and with current data some degree of uncertainty also comes with it. Nevertheless, an accurate treatment of the galaxy-related post-processing along the lines designed here, that hopefully will become feasible in the near future with more precise determinations of the SFR functions and of the enrichment history of galaxies at increasingly higher redshifts $z\ga 3$, will help in fully exploiting future GW observations and stellar evolution simulations to constrain the fundamental processes of stellar astrophysics that ultimately rule the formation and coalescence of binary compact remnants.

As a concluding remark, we point out that our approach can also be adapted with minimal change of formalism to multimessenger studies of various galaxy populations at different redshift. Most noticeably, it could be exploited to predict the rate of electromagnetic, neutrino, and cosmic ray emission events associated with NS-NS and/or BH-NS mergers as a function of host galaxy properties and of cosmic time, irrespective of detectability in the GW counterparts.

\begin{acknowledgements}
We acknowledge the referee for helpful comments. We warmly thank A. Bressan, F. Ricci, M. Spera, and J. Miller for stimulating discussions and critical reading. This work has been partially supported by PRIN MIUR 2015 “Cosmology and Fundamental Physics: illuminating the Dark Universe with Euclid”, and by the RADIOFOREGROUNDS grant (COMPET-05-2015, agreement number 687312) of the European Union Horizon 2020 research and innovation program. AL acknowledges the MIUR grant 'Finanziamento annuale individuale attivit\'a base di ricerca'.
\end{acknowledgements}

\newpage

\newpage

\newpage

\newpage

\newpage

\newpage

\newpage

\newpage

\newpage

\newpage

\end{document}